%% file: main.tex
\documentclass[trackchanges,twocolumn]{aastex701}

\newcommand{\m}{$\rm M_{\odot}$}
\newcommand{\mloss}{$\rm M_{\odot}\ yr^{-1}$}
\newcommand{\kms}{$\rm km\ s^{-1}$}
\newcommand{\s}{$\sim$}
\begin{document}

\title{Two years of shock interaction tracing three phases of evolution: the explosion of a Type IIn supernova, SN 2019vxm}

\shorttitle{SN~2019vxm: Long lived Type IIn SN}
\shortauthors{Gitika R. et al.}

\author[0009-0009-4872-1134]{Gitika Rameshan}
\affiliation{Indian Institute of Astrophysics, II Block, Koramangala, Bengaluru-560034, Karnataka, India}
\affiliation{Academy of Scientific and Innovative Research (AcSIR), Ghaziabad, Uttar Pradesh, 201002, India}
\email[show]{gitika.rameshan@iiap.res.in}

\author[0000-0002-0525-0872]{Rishabh Singh Teja}
\affiliation{Tsung-Dao Lee Institute, Shanghai Jiao Tong University, 1 Lisuo Road, Shanghai 201210, People’s Republic of China}
\email{rsteja@sjtu.edu.cn}

\author[0000-0002-6688-0800]{D. K. Sahu}
\affiliation{Indian Institute of Astrophysics, II Block, Koramangala, Bengaluru-560034, Karnataka, India}
\email{dks@iiap.res.in}

\author[0000-0003-3533-7183]{G. C. Anupama}
\affiliation{Indian Institute of Astrophysics, II Block, Koramangala, Bengaluru-560034, Karnataka, India}
\email{gca@iiap.res.in}

\author[0000-0001-9456-3709]{Masayuki Yamanaka}
\affiliation{Amanogawa Galaxy Astronomy Research Center (AGARC), Graduate School of Science and Engineering, Kagoshima University,
1-21-35 Korimoto, Kagoshima, Kagoshima 890-0065, Japan}
\email{yamanaka@sci.kagoshima-u.ac.jp}

\author[0000-0003-2611-7269]{Keiichi Maeda}
\affiliation{Department of Astronomy, Kyoto University, Kyoto, 606-8502, Japan}
\email{keiichi.maeda@kusastro.kyoto-u.ac.jp}

\author[]{Tatsuya Nakaoka}
\affiliation{Hiroshima Astrophysical Science Centre, Hiroshima University, 1-3-1 Kagamiyama, Higashi-Hiroshima, Hiroshima 739-8526, Japan}
\email{nakaoka@astro.hiroshima-u.ac.jp}

\author[]{Sota Goto}
\affiliation{Graduate School of Science and Engineering, Kagoshima University, 1-21-35 Korimoto,Kagoshima, Kagoshima 890-0065, Japan}
\email{k3110758@kadai.jp}

\author[0000-0001-7225-2475]{Brajesh Kumar}
\affiliation{South-Western Institute for Astronomy Research, Yunnan University, Kunming, Yunnan 650500, People's Republic of China}
\email{brajesh@ynu.edu.cn}

\author[0000-0003-2091-622X]{Avinash Singh}
\affiliation{The Oskar Klein Centre, Department of Astronomy, Stockholm University, AlbaNova, SE-10691 Stockholm, Sweden}
\email{avinash21292@gmail.com}

\author[0000-0002-4540-4928]{Miho Kawabata}
\affiliation{Okayama Observatory, Kyoto University, 3037-5 Honjo, Kamogatacho, Asakuchi, Okayama 719-0232, Japan}
\email{mkawabata@hep01.hepl.hiroshima-u.ac.jp}

\author[0000-0001-6099-9539]{Koji S Kawabata}
\affiliation{Hiroshima Astrophysical Science Center, Hiroshima University, Higashi-Hiroshima, Hiroshima 739-8526, Japan}
\email{kawabtkj@hiroshima-u.ac.jp}

\author[0000-0002-8482-8993]{Kenta Taguchi}
\affiliation{Department of Astronomy, Kyoto University, Kitashirakawa-Oiwake-cho, Sakyo-ku, Kyoto, Kyoto 606-8502, Japan}
\email{kentagch@kusastro.kyoto-u.ac.jp}

\begin{abstract}

We present multi-wavelength photometric and optical spectroscopic observations of the long-lived interacting supernova SN~2019vxm, spanning more than two years after the explosion. SN~2019vxm is a slowly rising (rise time \s 45.9 days in the \textit{R}-band), slowly declining supernova reaching an \textit{R}-band peak absolute magnitude of $\approx-$20.3 mag. The SN light curve post-maximum shows a shallow decline, followed by a secondary, steeper decline in the optical (0.01 mag day$^{-1}$), with late-time IR brightening. The total radiated luminosity is $5 \times 10^{50}$ erg, placing it among the energetic class of its type. We estimated a CSM mass of 3-8~\m\ through light-curve modeling (independent of the CSM density profile) and by comparison with theoretical models. We estimate a minimum ejecta mass of \s 3.88 \m\ from the broad H$\alpha$ component, consistent with the ejecta mass obtained from the light curve models. The solely interaction-dominated initial epochs are later accompanied by photon-scattering signatures, leading to asymmetric line profiles with symmetric wings. The late phase, characterized by enhanced brightness at longer wavelengths and a stronger asymmetric line profile with the red side flux strongly suppressed, indicates the influence of pre-existing or newly formed dust with temperatures \s 1500 K at \s $4\times 10^{16}$ cm. Even in the late phases, no nebular lines are present in the spectra, indicating dense or obscured ejecta.

\end{abstract}


\section{Introduction} 

Supernovae (SNe) are essentially the end of a star followed by an explosion. They are classified into thermonuclear and core-collapse (CC) SNe based on their explosion mechanisms. CCSNe occur when a star of mass, $M_{ZAMS}$ $\gtrsim$ 8 \m\ collapses due to gravity dominating over radiation pressure. The explosion that follows the collapse releases a large amount of energy and stellar material. They are primarily powered by radioactive Ni decay or H recombination in the envelope. SNe are classified into H-deficient (Type I) and H-rich (Type II) based on their optical spectral lines. There are further sub-classifications for these SNe based on several other spectral features. Type Ia, showing a prominent Si feature, are the thermonuclear class of SNe, while Type Ib and Ic are the He-rich and He-deficient CCSNe, respectively. The Type II SNe are also sub-classified based on their photometric and spectroscopic evolution \citep{1997ARA&A..35..309F}.

SNe can also be powered by the interaction of SN ejecta with the circumstellar material (CSM), termed interacting SNe. The CSM is sufficiently dense ($\rm \rho r^{2} \gtrsim 10^{14}\ g\ cm^{-1}$); this can convert a substantial fraction ($\lesssim 50\%$) of the ejecta kinetic energy into radiation \citep{2017hsn..book..403S,2014ARA&A..52..487S}. The slow-moving CSM, ionized primarily by UV and X-ray photons, produces narrow emission lines in the spectrum. Type IIn was the first observed subtype of this class, where `narrow' Balmer features in the optical spectra were observed due to interaction of ejected material with H-rich CSM \citep{1990MNRAS.244..269S}. Other relatively rare classes include Type Ibn and Icn, in which the progenitors are stripped of their outer envelopes and have He- and C/O-rich CSM \citep{2017hsn..book..403S, 2020RSOS....700467F}. There are other classes of transients, such as Fast Blue Optical Transients (FBOTs), whose observational properties can also be explained by such interactions \citep{2019MNRAS.488.3772F,2022ApJ...933..238M,2024ApJ...972..140K}.

Due to the efficient energy conversion, interacting SNe are very luminous, reaching a mean \textit{R/r}-band magnitude of \s -18.9 mag \citep{2020A&A...637A..73N}. Type IIn SNe can be broadly classified into fast and slow risers/decliners depending on their \textit{R}-band light curve evolution \citep{2020A&A...637A..73N,2025ApJ...987...13R,2024arXiv241107287H}.
These SNe also exhibit significant diversity in their spectroscopic properties. The spectrum is dominated by three components: a slow-moving CSM with v$\rm\sim100\ km\ s^{-1}$, the cold dense shell (CDS), sweeping up the CSM as it propagates with v$\rm\sim1000\ km\ s^{-1}$, and a fast-moving ejecta with v$\rm\sim10000\ km\ s^{-1}$ \citep{1997Ap&SS.252..225C}. The spectra observed with higher resolution also show narrow absorption features from the slow-moving CSM in the form of P-Cygni profiles, as seen in SN 2010jl \citep{2014ApJ...797..118F}. The P-Cygni profile from ejecta could also appear in the later phases \citep{2024MNRAS.530..405S}. The line emission could be a combination of double-peaked or complex multi-component profiles, which, again, probe the intricacies of this class of objects. Asymmetries in line profiles are attributed mainly to asymmetry in the CSM or ejecta, the presence of pre-existing dust, and occultation by the receding photosphere or CDS \citep{2017MNRAS.471.4047A,2025ApJ...983..101B,2011ApJ...732...63S,2024ApJ...977..152F}. The intrinsic ejecta/CSM asymmetry is usually validated by polarization measurements, while dust is accompanied by an IR excess \citep{2025A&A...702A.213R}.

This interesting and intriguing class of transients needs progenitors with elevated mass loss rates ranging between $\dot{M}\sim 10^{-3}-1.0$ \mloss \citep{2014ARA&A..52..487S}, not completely achieved by classical stellar winds \citep[$\dot{M}\sim 10^{-6}-10^{-3}$ \mloss;][]{1999isw..book.....L}. Massive stars like Wolf-Rayet stars and red/yellow supergiants can also undergo enhanced mass loss through nuclear flashes, pulsational pair instability, or convection waves absorbed by the envelope \citep{2012MNRAS.423L..92Q, 2014ARA&A..52..487S,2024arXiv240504259D}. Luminous blue variable star is a potential progenitor supported by direct detection, like that of SN 2005gl \citep{2009Natur.458..865G} or by precursor eruptions like SN 2009ip \citep{2011ApJ...732...32F,2013ApJ...767....1P,2013MNRAS.430.1801M}. In addition to massive stars, stars in binary systems can also lead to such high mass-loss rates. Mass transfer occurs in almost 70$\%$ of all massive stars \citep{2012Sci...337..444S}, and the observed fraction of SNe agrees with recent simulations \citep{2026A&A...706A.169E}.
 
SNe IIn exhibit a myriad of evolutionary profiles, with diversity in luminosity, dynamics, interactions, CSM morphology, and dust formation. Thus, a comprehensive understanding is only possible if we have a very large sample space of well-studied events that incorporates all the diverse evolutionary trends. The rarity of these events \citep[$\sim 5\%$ among CCSNe;][]{2023A&A...670A..48C} further makes it necessary to observe as many such events as possible to understand the evolution of their progenitors and the mass loss mechanisms.

SN 2019vxm (ASASSN-19acc) was discovered on 2019 December 01 (JD 2458818.5) at $\alpha$=19:58:28.540, $\delta$= +62:08:15.83 by the ASAS-SN survey \citep{2019TNSTR2492....1S}. It was then classified as a Type IIn SN at a redshift of z=0.019 \citep{2019TNSCR2506....1L}. It has a diffuse galaxy neighbor, SDSS J195828.83+620824.3, although it remains unclear whether that is the host \citep{2026arXiv260523637L}. 
SN~2019vxm has been widely followed and has been included in several recent studies. \cite{2024AN....34530166T} presented the optical photometric evolution until \s 781 days from discovery and spectra at two epochs. \cite{Lane_2026} discussed the spatial and temporal association of SN 2019vxm with GRB 191117A with a 3.3$\sigma$ confidence. They model the light curve up to 200 days, after which they concluded that the progenitor was a massive star of \s 40 \m\ that had lost around \s 1.5 \m\ of mass, forming a massive CSM. \cite{2026arXiv260523637L} modeled the photometric observations using multiple interaction and Ni decay models for both optically thick and optically thin CSM, favoring the latter. They also conclude that the pre-existing dust is responsible for the IR excess.

In this paper, we present UVOIR photometric and optical spectroscopic observations of SN 2019vxm. The structure of the paper is as follows: We provide the details of the data obtained and the reduction procedure in Section \ref{obs_data}, followed by the distance and extinction estimation in Section \ref{dist}. We discuss the multiband photometric evolution and colors in Section \ref{phot} and also compare them with other well-studied long-duration interacting SNe. We also model the SED and discuss its evolution. In Section~\ref{model}, we perform the light curve modeling to constrain ejecta and CSM properties. This is followed by inferences from the optical spectra in Section \ref{spec}. Lastly, in Section~\ref{discuss_sec}, we examine the light curve evolution and line asymmetries, and discuss possible scenarios that can explain the observed properties.
 
\section{Observations and Data Analysis} \label{obs_data}

\begin{figure}
    \centering
    \includegraphics[width=1.0\linewidth]{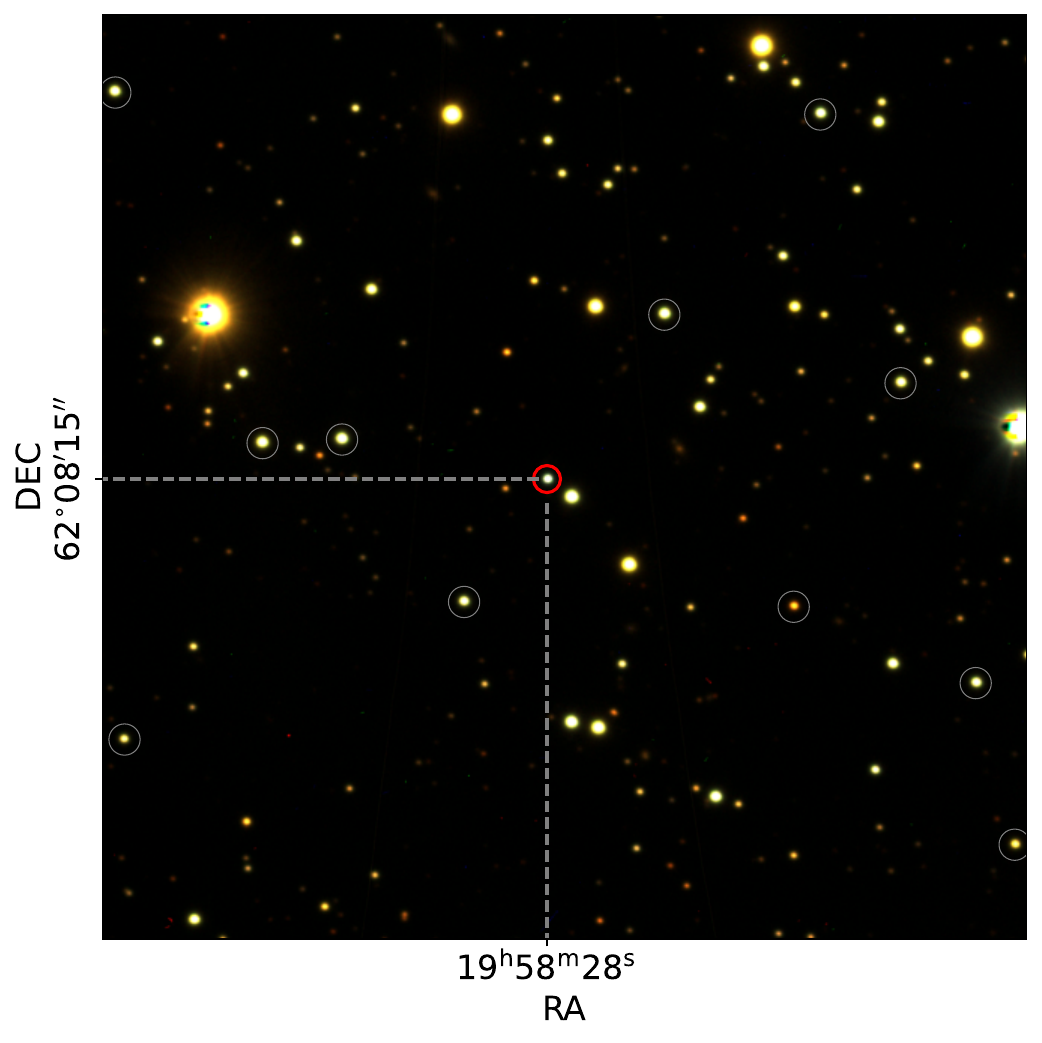}
    \caption{A color composite image of SN~2019vxm in \textit{B, V, R} filters at t \s 20 days (2012 December 07). The red circle corresponds to SN~2019vxm. The white circles mark the secondary calibrators used for photometry.}
    \label{color_image}
\end{figure}

Optical observations of SN 2019vxm were primarily obtained using the 2-m Himalayan Chandra Telescope (HCT) located at the Indian Astronomical Observatory \citep[IAO;][]{2014PINSA..80..887P}. We also obtained the Near Infrared (NIR) photometric observations using the Kanata Telescope. Figure \ref{color_image} shows the color composite image marking the SN and secondary stars used for photometric calibrations. In addition to optical and NIR, publicly available data from the \textit{Swift} Ultraviolet/Optical Telescope \citep[UVOT,][]{2004ApJ...611.1005G,2005SSRv..120...95R} were used for the analysis. Pre-explosion data from the Asteroid Terrestrial-impact Last Alert System \citep[ATLAS,][]{2018PASP..130f4505T} were used to search for any mass-loss events prior to the SN discovery.

\subsection{Photometric Observations}

\subsubsection{Optical Photometry}
Optical photometric monitoring campaign for SN 2019vxm ran from 2019 December 3 to 2021 November 1 using Hanle Faint Object Spectrograph Camera (HFOSC) mounted on the HCT. The Bessel-\textit{U, B, V, R, I }filters were used for observations.

The science frames were corrected for bias using master bias frames and flat-fielded, with master flat frames created from multiple twilight- or dawn-exposures in each filter. On photometric nights, along with SN frames, standard star frames from \citet{1992AJ....104..340L} were also observed. Aperture photometry of Landolt's standard stars and the stars (marked in Figure \ref {color_image}) in the supernova frame was performed at an optimal aperture. The stars in the supernova frame were calibrated using the average color term of HFOSC and zero points determined using Landolt standard stars. These secondary standard stars in the supernova frame were later used for calibrating supernova magnitudes to the standard system. On all other nights, the instrumental magnitude of the supernova and the secondary standard stars was obtained from point-spread-function (PSF) photometry. An aperture correction was applied to account for the PSF wings.

At late phases, when the SN became very faint, multiple frames were obtained and combined to increase the signal-to-noise ratio. Aperture photometry at FWHM (obtained from the bright stars in the frame), followed by aperture correction, was used to obtain the SN magnitude. 

 \subsubsection{NIR Photometry}
The near-infrared (NIR) imaging observations were carried out on 55 nights from 2019 December 4 to 2022 July 31 using the Hiroshima Optical and Near-InfraRed camera \citep[HONIR;][]{2014SPIE.9147E..4OA} attached to the 1.5-m Kanata telescope. Dark subtraction and flat-field correction were performed in the standard manner. After that, the sky background was subtracted. PSF photometry was carried out for the SN and the local standards. The photometric calibration was performed using the same field stars as the object from the 2MASS catalog \citep{2003yCat.2246....0C}.

\subsubsection{UV Photometry}
We have also included the early-phase \textit{Swift/}UVOT observations from the open-source Swift archives\footnote{\url{https://www.swift.ac.uk}}. We used a custom Python script to perform photometry using HEASOFT tools and the latest calibration database for the UVOT instrument. We defined a 5 arcsec aperture for the SN and a similar nearby region for the background. We obtained SN photometry in the \textit{UVW2}, \textit{UVM2}, and \textit{UVW1} filters in the near-UV regime, and in the \textit{U}, \textit{B}, and \textit{V} filters in the optical.

All the magnitudes are given in the Vega magnitude system and tabulated in Appendix~\ref{append}.

\begin{figure*}
    \centering
    \includegraphics[width=1.0\linewidth]{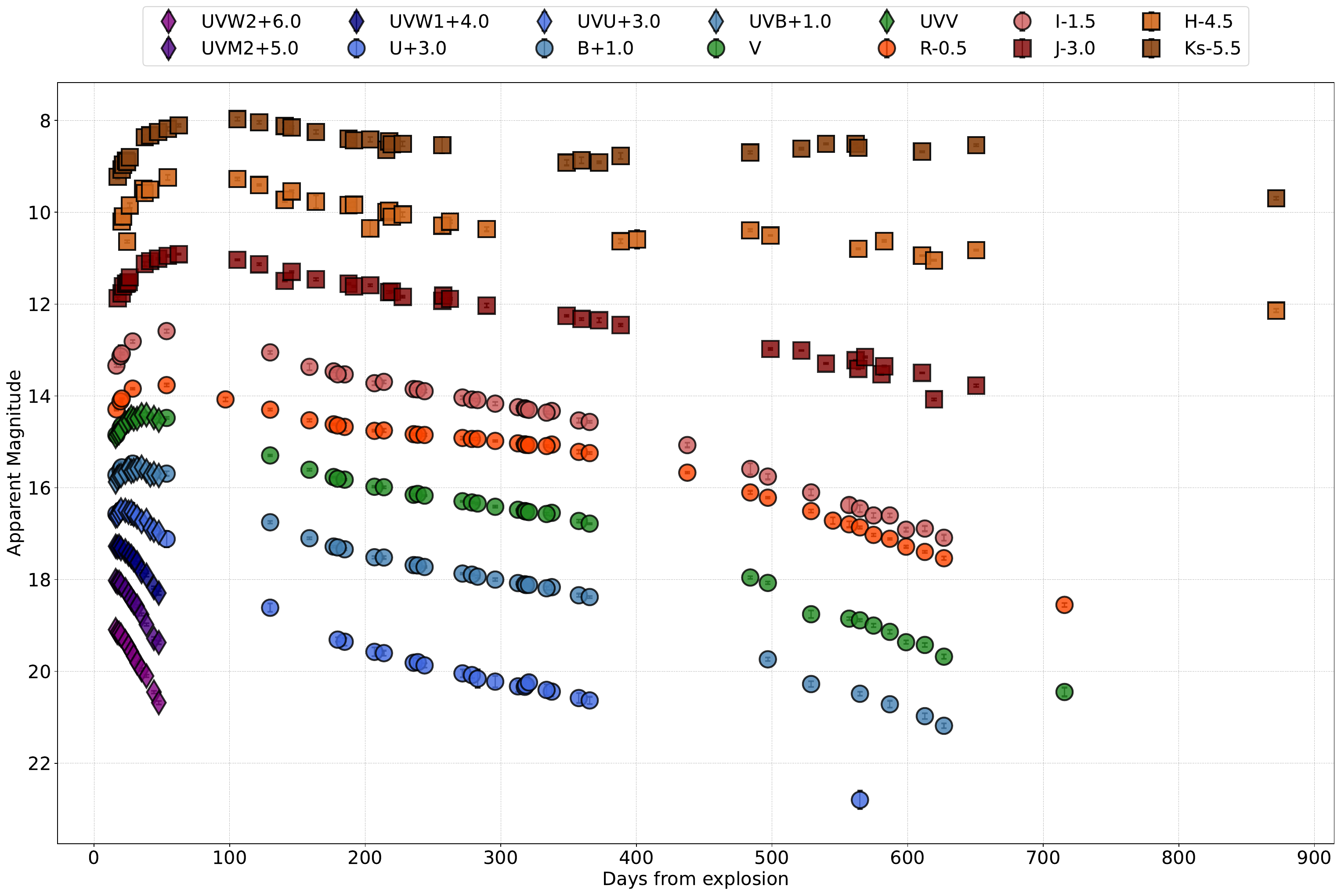}
    \caption{Light curve of SN 2019vxm from UV to the NIR obtained using HCT (circular markers), Kanata (square markers), and Swift (diamond markers) facilities. (Constant offsets are applied for visual clarity.)}
    \label{LC}
\end{figure*}

\subsection{Spectroscopic Observations}

The spectroscopic follow-up observations with HCT were performed from 2019 December 3 to 2021 November 1 using two grisms, Gr7 (3500–7800 \AA) and Gr8 (5200–9250 \AA), with spectral resolutions of 1000 and 1200, respectively. The spectroscopic data were reduced in a standard manner using the appropriate IRAF tasks \citep{1986SPIE..627..733T}. The obtained spectra in pixel coordinates were converted to the wavelength scale using arc lamp spectra of FeAr and FeNe. The instrument's wavelength-dependent response function was obtained from spectra of the spectrophotometric standard observed on the same night. The spectra from both grisms were combined if obtained on the same night. To account for photon loss, these fluxes were further calibrated using multi-band photometric data. 

\section{Distance and Extinction} \label{dist}
The magnitudes are corrected only for Milky Way extinction since the spectra showed no distinct \ion{Na}{1D} lines from the host (also reported by \cite{2024AN....34530166T}). We use an extinction of $A_{V} = 0.282$ obtained using \cite{2011ApJ...737..103S} via NASA/IPAC. We use the extinction law by \cite{2007ApJ...663..320F} assuming $R_{V}=3.1$ through the \textit{extinction} module of \textit{python} for extinction correction across all photometric bands and spectra. While \cite{Lane_2026} included galactic extinction after modeling the host galaxy SED, \cite{2026arXiv260523637L} discuss the uncertainty associated with the galaxy being the host due to a questionable redshift measurement of the galaxy.

The reported SN redshift, $z\sim0.019$, is used throughout. We use the cosmological model with $\rm H_{0}=73\ km\ s^{-1} Mpc^{-1},\ \Omega_{M}=0.27\ and\ \Omega_{\lambda}=0.73$ to obtain Virgo infall corrected distance modulus of 34.58 $\pm$ 0.15 mag\footnote{\url{http://ned.ipac.caltech.edu/}} for our calculations. We consider an explosion epoch of JD 2458804.53 $\pm$ 0.30 throughout the paper obtained from the extensive TESS \textit{R}-band coverage at rise given in \cite{Lane_2026}. We use this as the reference time throughout our work.

\section{Photometric Analysis} \label{phot}

\subsection{Light Curve} \label{lc_sec}

The time evolution of UV, optical, and NIR light curves is given in Figure~\ref{LC}. The SN exhibits a rapid decline in the UV regime (observations until t \s 48 days), while the optical evolution is slow, with observations until 716 days after explosion. The SN light curve reaches a maximum, which is not observed in \textit{UVM2, UVW1, or UVW2} due to a lack of early data and shorter rise-time to maximum. The redder bands rise sequentially to their maxima, with longer rise times as the supernova cools. Using the explosion epoch estimated by \cite{Lane_2026} the rise time using spline fit in \textit{U, B, V, R} and \textit{I} bands are \s 22.8, 25.3, 38.8, 45.9, and 62.6 days, respectively, and in \textit{J, H, K} bands are 66.9, 70.7, and 96.4 days, respectively.

The \textit{R}-band rise time of \s 45.9 days places it at the lower end of slow risers. The peak \textit{R}-band absolute magnitude is \s -20.3 mag. The SN then decays slowly, with a shallow, plateau-like decline, possibly due to interaction, until 400 days, after which it decays faster. The decline rate until the first 50 days from maximum is \s 0.7 mag (100~d)$^{-1}$. Thus, it falls in the category of slow decliners (taken as $\rm\Delta t \sim 1.3\ mag\ (100~d)^{-1}$ in \cite{2020A&A...637A..73N}) and decays by 1 mag in \s 210 days from peak. The shallow decay, using a simple linear fit from t \s 180-350 days, yields decay rates of 0.67 $\pm$ 0.03, 0.54 $\pm$ 0.01, 0.48 $\pm$ 0.01, 0.27 $\pm$ 0.01, and 0.53 $\pm$ 0.01 per 100 days in the \textit{U, B, V, R,} and \textit{I}-bands, respectively. The decay rates in \textit{B, V, R,} and \textit{I}-bands at t$>$ 400 days are \s 1.03-1.07 mag/100 days.

\subsection{UVOIR bolometric luminosity}

\begin{figure}[hbt]
    \centering
    \includegraphics[width=1.0\linewidth]{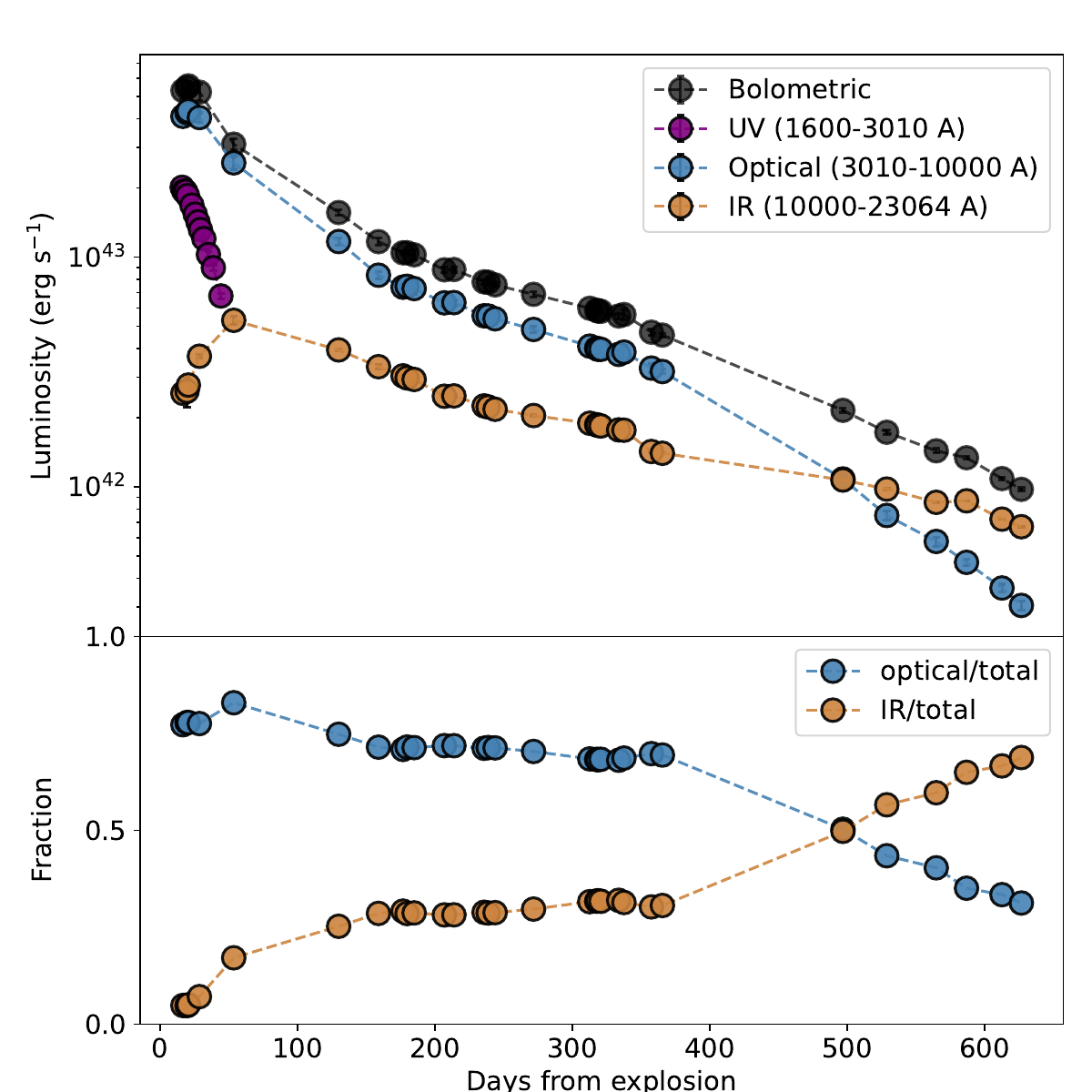}
    \caption{\textbf{Top:} The evolution of the pseudo-bolometric luminosity (1600-23064\AA) along with the UV (1600-3010 \AA), optical (3010-10000 \AA), and IR (10000-23064  \AA) contribution with time. \textbf{Bottom:} The fraction of optical and IR luminosity compared with the pseudo-bolometric luminosity. The clear crossover from an optical to an IR-dominated luminosity is visible in both panels. }
    \label{all_lum}
\end{figure}

The pseudo-bolometric luminosity was calculated by integrating the UVOIR flux in the range 1600-23064~\AA\ using the trapezoidal method. The luminosity contributions from the UV, optical, and IR bands were calculated by integrating the flux over the wavelength ranges 1600-3010~\AA, 3010-10000~\AA, and 10000-23064~\AA, respectively. Figure \ref{all_lum} represents the evolution of the pseudo-bolometric luminosities with time as well as the optical and IR fraction contributing to the pseudo-bolometric luminosity.
We have Swift/UVOT observations only until t $\sim$ 48 days. However, we note that the UV contribution, which is $\sim 37\%$ at t \s 16 days, reduces to $\sim 18\%$ at 44 days. Thus, even though UV flux is not available beyond t\s 50~days, the contribution is not significant. Only sparse \textit{U} band observation is present at the late phase (t $>400$ days) compared to the other bands. Hence, the \textit{U}-band magnitude at a later phase (t$>$400 d) was calculated by first quantifying its contribution to the optical by comparing fluxes with and without the \textit{U}-band in one epoch at the late phase where we have observations, and then applying this scaling factor to the BVRI luminosity. The flux with \textit{U}-band and without \textit{U}-band maintained a ratio of 1.07 from t\s 337 to 364 days, while it was 1.06 at t\s 564 days, and we have used the latter for calculations of luminosity. Integrating the obtained bolometric luminosity until t\s 612 days gives a total radiated luminosity of $\rm E_{rad}\sim 5.23\times 10^{50}$ erg. The contributions of different bands to the bolometric luminosity indicate that the IR contribution increases from $\sim 5\%$ at t\s 16 days to $\sim 29-32\%$ at 158-317 days as it reaches the plateau, and after t \s 450 days the IR contribution rises to $>69\%$ at t \s 629 days. On the other hand, the contribution of optical luminosity to bolometric luminosity initially declines from $\sim 77\%$ to $\sim 69 \%$ as it evolves from t \s 16 days to 365.5 days, and contributes only $\sim 31\% $ at a later phase (t\s 612 days). The UV flux contributes $\sim 36\%$ to the pseudo-bolometric luminosity near the maximum. The bolometric luminosity shows three decline trends: a fall from maximum to \s 170 days, followed by a shallow decline until \s 365 days, and then a steeper decline. The decline rates are
1.21 $\pm$ 0.08 mag 100d$^{-1}$, 0.44 $\pm$ 0.01 mag 100d$^{-1}$ and 0.63 $\pm$ 0.03 mag 100d$^{-1}$ for each of these phases respectively. 

\subsection{Comparison with other IIns}

\begin{figure}[hbt!]
    \centering
     \resizebox{\hsize}{!}{\includegraphics{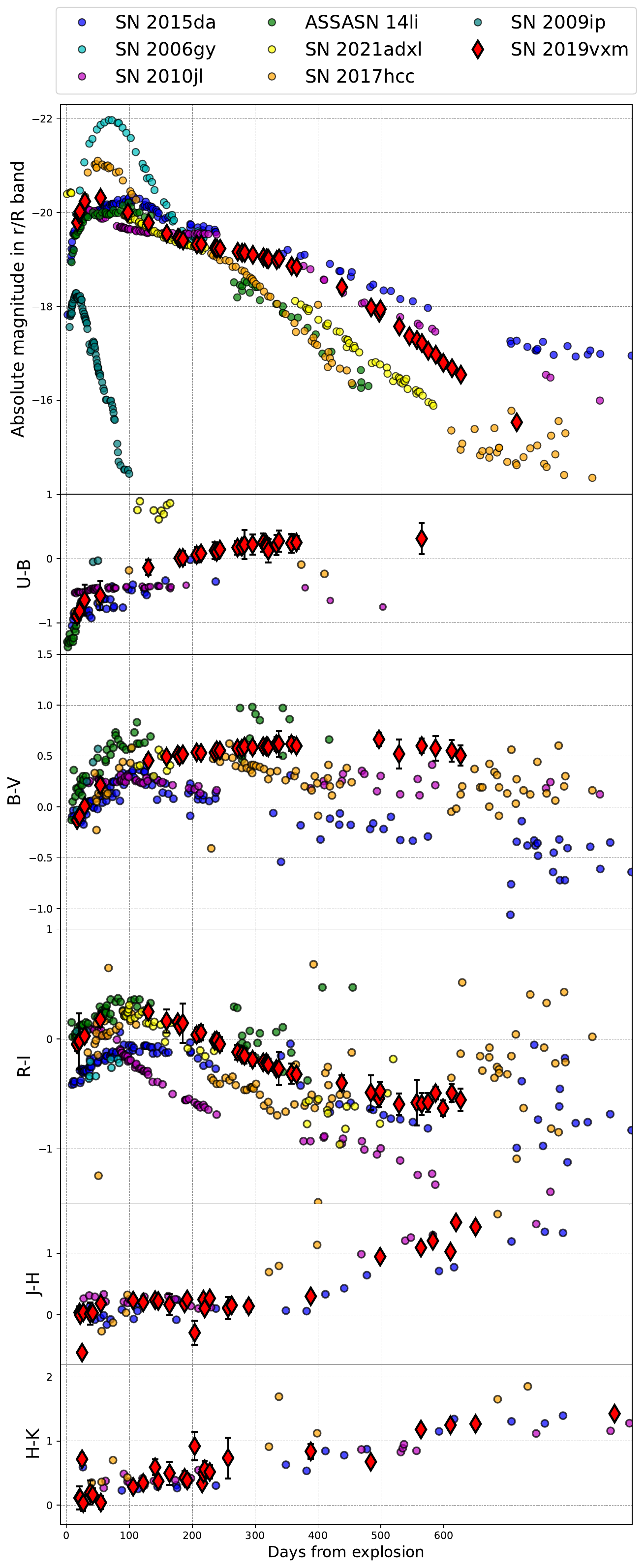}}
    \caption{\textbf{Top:} Comparison of \textit{R/r}-bandlight curve of different SNe IIn, including an SLSN, SN 2006gy. \textbf{Bottom:} \textit{U-B, B-V, R-I, J-H,} and \textit{H-K} color evolution of SN 2019vxm and comparison with colors of the sample set of SNe IIn.}
    \label{lc_comp}
\end{figure}

Even though SNe IIn are broadly classified as short- and long-lived, the objects in each class exhibit a wide diversity in their light curves. Figure \ref{lc_comp} represents the evolution of SN 2019vxm in comparison with other IIn SNe in the \textit{R/r}-band. SN 2015da has a peak magnitude similar to SN 2019vxm, but longer rise time ($\gtrsim$ 100 days) with an estimated CSM mass of \s 20 \m \citep{2020AA...635A..39T,2024MNRAS.530..405S}. It also decays more slowly than SN 2019vxm. The same initial evolution is also observed in ASASSN-14il \citep{Dukiya_2024}, even though it decays faster than both SN 2015da and SN 2019vxm. It has an estimated CSM mass of 4.7-9.1 \m, which is concluded to have been lost through eruptive mass loss (\citeauthor{Dukiya_2024}). On the other hand, SN 2006gy is an SLSN with a peak magnitude of $M_{\text{R}}\sim -22$~mag and a light curve evolution that is very different from SN 2019vxm in its rise, maximum, and decline \citep{2007ApJ...666.1116S}. SN 2017hcc is brighter than SN 2019vxm, with a slightly longer rise time but with a faster decline \citep{2023AA...669A..51M,2026ApJ..1001..169S}. SN 2021adxl and SN 2010jl are SNe that were discovered after they reached their peak brightness, and the rise time and maximum magnitude phase went unobserved due to solar conjunction \citep{2024AA...690A.259B, 2012AJ....144..131Z}. However, the light curve of SN 2010jl also shows two distinct post-maximum phases: a plateau and a steeper decline, and the CSM mass is estimated to be $\gtrsim$ 3 \m \citep{2014ApJ...797..118F}. SN 2019vxm, even though following a different light curve evolution, is comparable to the rest of the long-lived IIn SNe. The evolution, however, is in stark contrast to that of the short-lived SN 2009ip. SN 2009ip is about 2 magnitudes fainter than SN 2019vxm and drops by nearly 4 magnitudes in 100 days. This indicates that SN 2019vxm is in the stronger-interacting regime.

\begin{figure*}[hbt!]
    \includegraphics[width=0.33\linewidth]{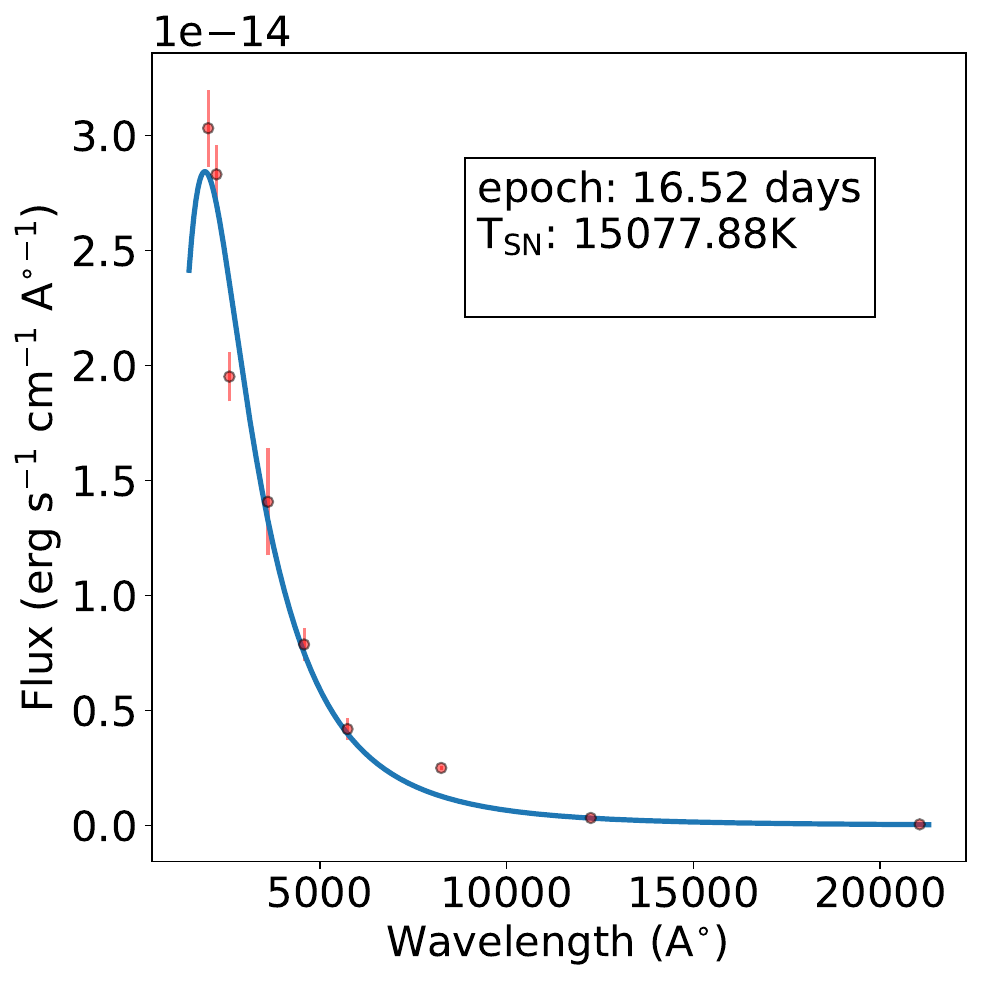}
    \includegraphics[width=0.33\linewidth,trim={0cm 0cm 1.5cm 1.0cm},clip]{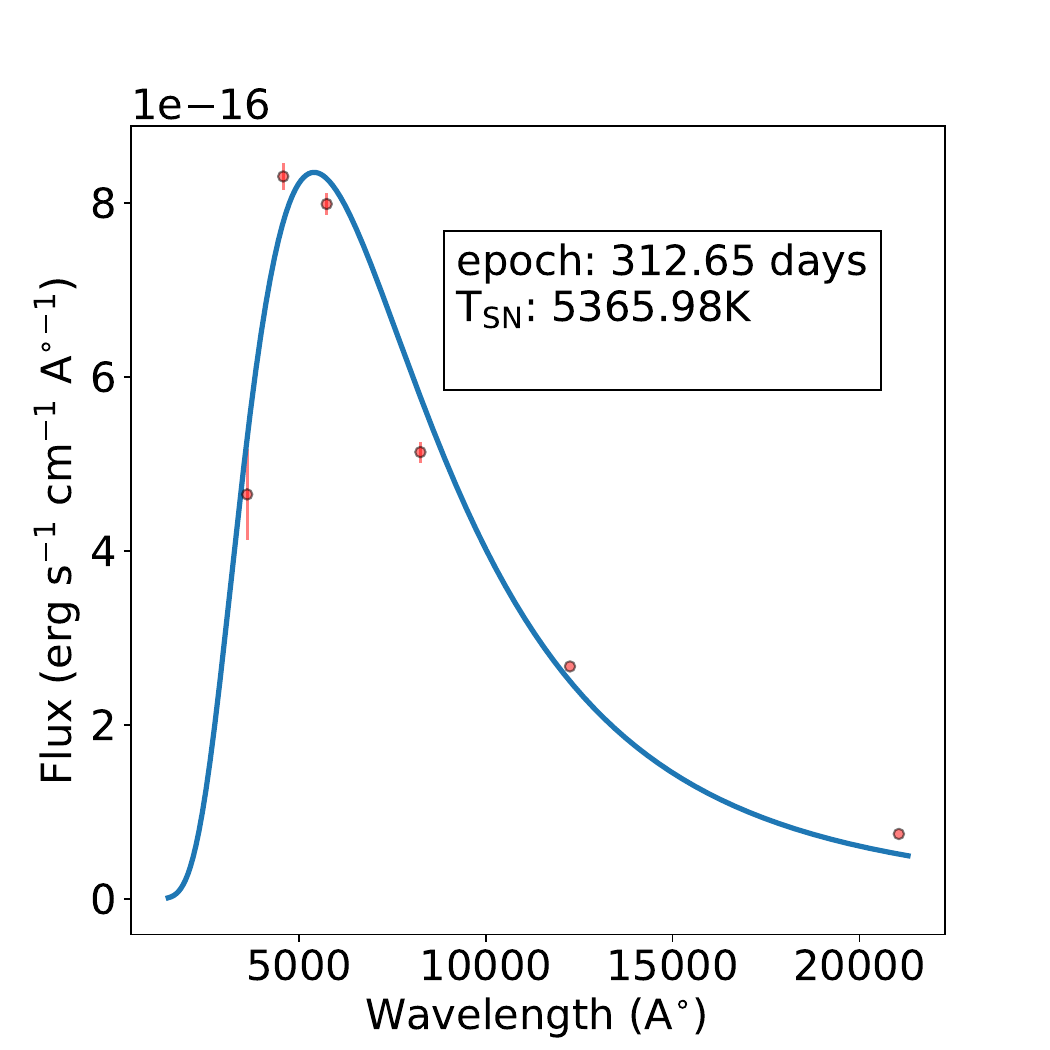}
    \includegraphics[width=0.33\linewidth]{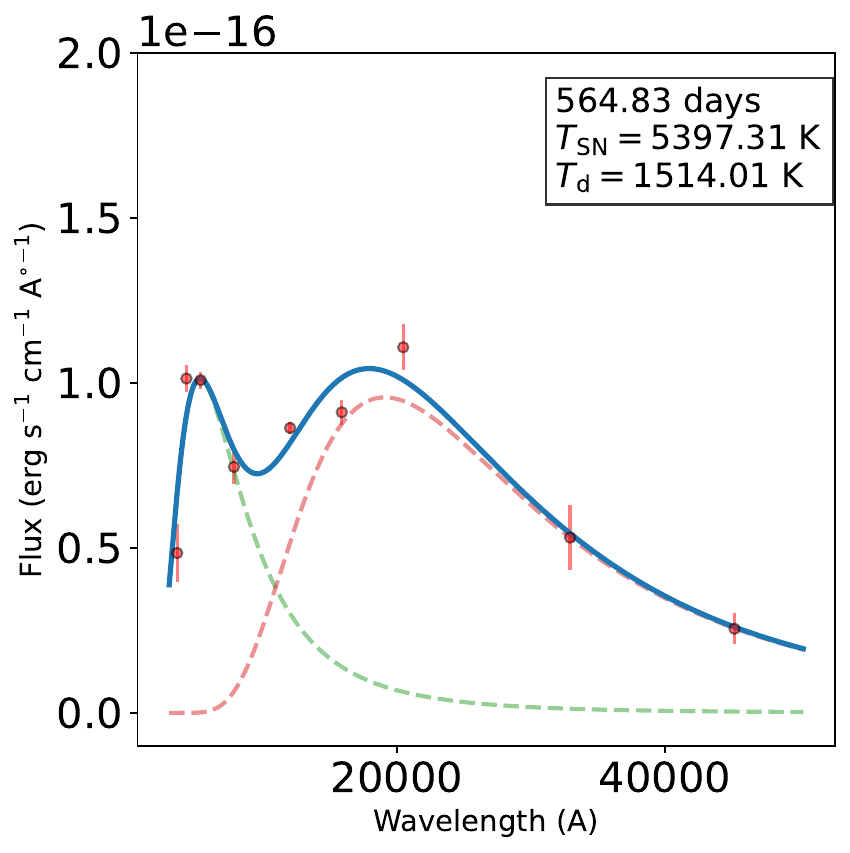}
    \caption{The evolution of the SEDs with time, where the red data points are the flux densities from photometric observation and the blue is the best fit model. Single blackbody fit for two epochs, t \s 16.5 days and t \s 312.6 days, and a dual blackbody fit for one epoch, t \s 564.8 days, along with the SN and the dust temperature specified.}
    \label{SED}
\end{figure*}

We also compare the colors of SN 2019vxm with other Type IIn SNe in the lower panels of Figure \ref{lc_comp}. The densely sampled \textit{U}-band light curve of SN 2019vxm enabled us to observe \textit{U-B} increases over time; the \textit{U-B} color of SN 2019vxm becomes progressively redder than that of SN 2010jl. The \textit{B-V} color is more useful for comparison with the sample, since \textit{U}-band observations for other SNe are sparse at later epochs. The \textit{B-V} color for all SNe in the sample is similar for the first 100 days. The \textit{B-V} color of SN 2019vxm increasingly becomes redder than SN 2010jl and SN 2015da, after which it settles down to a constant color from t \s 300 days. However, SN 2019vxm has a lower \textit{B-V} than ASASSN-14il in the initial epochs. It appears to trace the \textit{R-I} color in the same manner, lying at the higher end of the sample, but with an evolution similar to that of SN 2017hcc, SN 2015da, and SN 2021adxl. All these SNe initially rise, and at \s 200 days, the SN starts being brighter in the \textit{R}-band. This phase is reached much earlier in SN 2010jl, whereas ASASSN-14il does not exhibit this phase.
The \textit{J-H} and \textit{H-Ks} for all SNe in the sample show a similar slow rise, and SN 2019vxm is no exception. The \textit{J-H} color remains near zero until t \s 300 days, after which it steadily increases while the \textit{H-Ks} color rises from the initial phases. However, we also note that the sample presented here is very limited, with only a subset of Type IIn SNe observed in the IR bands.

 \subsection{SED modeling}

We obtained the spectral energy distribution (SED) at epochs with simultaneous multiband observations (including UV bands up to t \s 48 days), excluding the \textit{R} band since it is dominated by the strong H$\alpha$ emission. We used UVOIR observations to model these SEDs, interpolating the JHK LCs. Until t \s 400 days, all the points were fit by a single black body. However, at a later time (t $>$ 400 days), the SED showed an IR excess that could not be fit by a single blackbody. We fit the \textit{JHK}-bands and the Wise \textit{W1, W2}-bands from \cite{2026arXiv260523637L} at these epochs with another blackbody, which corresponds to optically thick dust. We use JHK points observed nearest to the optical observation, within a $\pm$ 3-day window (with the t\s 496-day IR point relaxed to a 5-day window), and for the Wise bands, we linearly interpolate the IR LC to obtain magnitudes corresponding to these epochs. Thereafter, we model the SED using the relation:
\begin{equation}
    F_{\lambda}(\lambda)=\frac{{R_{SN}}^{2}}{{D_{L}}^{2}}B_{\lambda}(\lambda,T_{SN})+\frac{{R_{d}}^{2}}{{D_{L}}^{2}}B_{\lambda}(\lambda,T_{d})
\end{equation}
where only the first term is used for modeling up to t \s 400 days. Here, $T_{SN}$ and $R_{SN}$ are the photospheric temperature and radius, while $T_{d}$ and $R_{d}$ are the temperature and radius of the IR emitting region. $B_{\lambda}(\lambda,T)$ is the Planck black body function in units of $\rm erg\ s^{-1}\ cm^{-2}$ \AA$^{-1}$. $D_{L}$ is the distance to the supernova adopted as 82.6 Mpc from NED.

The resulting best-fit SEDs for some of the epochs are shown in Figure \ref{SED}. The blackbody temperature corresponding to the photosphere drops drastically from $>$ 15000 K at t \s 16.5 days to \s 5500 K. This is also accompanied by a rise in the blackbody radius from \s $2\times 10^{15}$ cm to $7.5\times 10^{15}$ cm. This high-temperature blackbody traces the photosphere of the SN, which expanded outward until t \s 130 days and then started receding inwards. The temperature of the receding photosphere is constant throughout (\s 5000 K). The temperature and radius of the NIR emitting region, including Wise W1, W2 data, are tabulated in Table \ref{dust_table}. The dust temperature estimated is \s 1600 K. The radius of the NIR emitting region is \s $4 \times 10^{16}$ cm, which is the lower limit since we are assuming the dust to be optically thick.

\begin{deluxetable}{cccc}[h]
\label{dust_table}
\tablewidth{1\linewidth} 
\tablecaption{Best-fit dust temperature and radius}
\tablehead{
\colhead{Epoch (days)} & \colhead{$T_d$ (K)} &  \colhead{$R_d$ ($\times 10^{16}$ cm)} 
}
\startdata
483.97 & $1650.04^{+382.89}_{-365.74}$ & $4.0^{+0.01}_{-0.01}$\\
& & \\
496.91 & $1877.44^{+320.5}_{-361.0}$ & $3.02^{+0.01}_{-0.01}$ \\
& & \\
556.89 & $1691.54^{+290.67}_{-431.26}$ & $3.79^{+0.01}_{-0.01}$ \\
& & \\
564.83 & $1510.25^{+82.45}_{-82.68}$ & $4.25^{+0.61}_{-0.54}$ \\
& & \\
586.87 & $1615.05^{+306.69}_{-256.97}$ & $3.91^{+0.01}_{-0.01}$ \\
& & \\
612.69 & $1660.27^{+383.91}_{-347.65}$ & $4.0^{+0.01}_{-0.01}$ \\
& & \\
871.25 & $1215.52^{+365.23}_{-372.41}$ & $4.2^{+0.01}_{-0.01}$\\
\enddata
\end{deluxetable}

\section{Light curve Modeling} \label{model}

We modeled the UV-optical light curve of SN 2019vxm using MOSFiT \citep{2018ApJS..236....6G}, a Python-based open source code for fitting transient light curves based on Bayesian inference. It has built-in models that incorporate radioactive Ni decay (\textit{default}), interaction with csm (\textit{csm model}), interaction with Ni decay (\textit{csmni model}), magnetar-powered model (\textit{magnetar model}), among others. The interaction model accounts for the conversion of ejecta KE into radiation through shock heating of the ambient CSM \citep{2012ApJ...746..121C,2013ApJ...773...76C}. The density of the ejecta and CSM follows a power law profile as the ejecta expands uniformly. \cite{2025ApJ...987...13R} studied IIn SNe population properties using parameters derived from MOSFiT.

\begin{figure*}[htb!]
    \centering
    \includegraphics[width=1.0\linewidth,trim={7cm 0cm 7cm 0.0cm},clip]{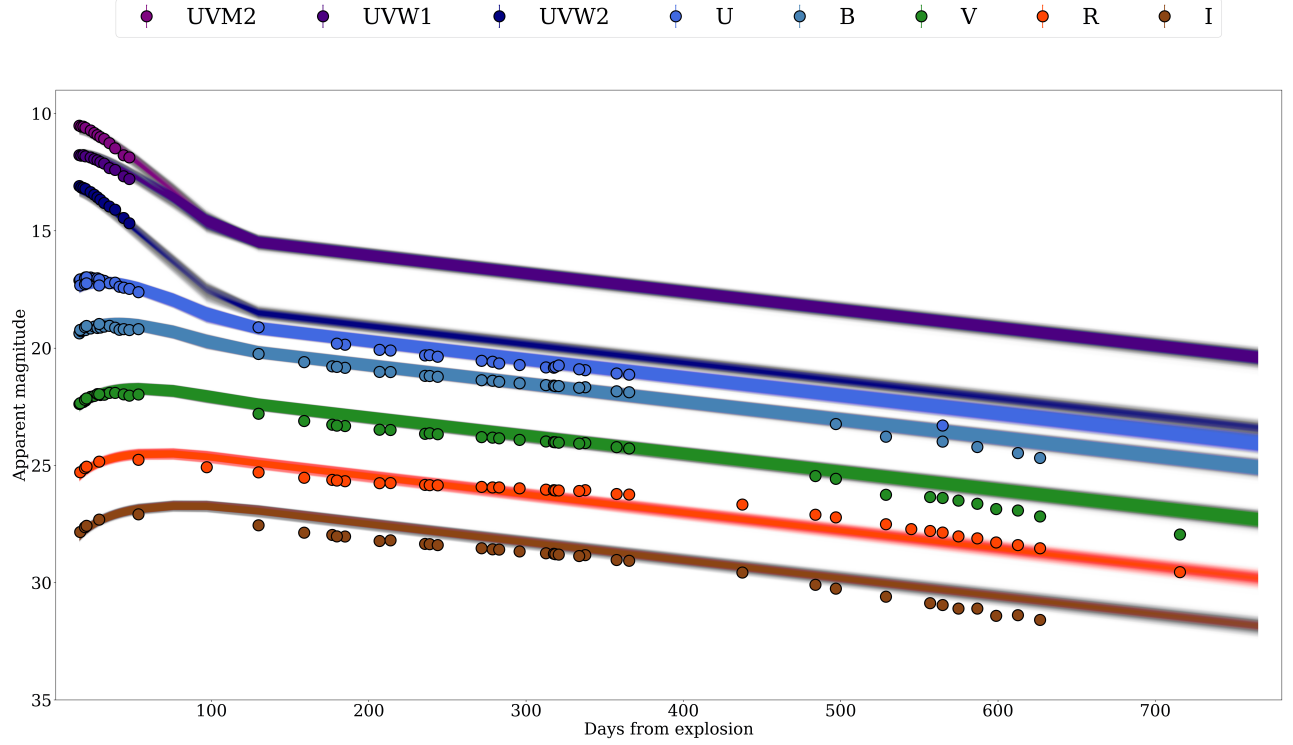}
    \caption{The converged fit obtained for LC up to t \s 400 days for a fixed $s=2.0$. The points correspond to the data, while the fit is the model with posterior distribution (and thus the errors) representing the thickness. The model LC has been extended to t\s 750 days for comparison.}
    \label{mosfit_lc}
\end{figure*}

\begin{table*}[hbt!]
\centering
    \caption{The parameters obtained by modeling the light curve by using \textit{csm} model of MOSFiT.}
    \resizebox{1.0\textwidth}{!}{
    \begin{tabular}{c c c c c c c c c c}
    \hline
    Duration & s & $\rm M_{CSM}$ & $\rm M_{ejecta}$ & n & $\rm n_{H}$ & $\rm R_{0}$ & $\rm \rho$ & $\rm T_{min}$ & $\rm v_{ejecta}$ \\
    (days) & &  \m &  \m & & $(\times 10^{17} cm^{-3})$ & (AU) &  $(\times 10^{-12} g\ cm^{-3})$ & (K) & $\rm (km\ s^{-1})$\\
    \hline
    \hline
    & & & & & & & & &\\
    200 & 1.4$^\dagger$ & $3.16^{+1.20}_{-1.02}$ & $7.41^{+6.08}_{-2.28}$ & $8.55^{+1.2}_{42}$ & $2.34^{+15.85}_{-1.62}$ & $37.15^{+17.80}_{-26.92}$ & $14.45^{+35.66}_{-2.97}$ & $6025.60^{+140.35}_{-137.16}$ & $8222.42^{+133.60}_{-94.12}$\\
    & & & & & & & & &\\
    400  & 0.55$^{+0.51}_{-0.36}$ & $5.01^{+1.30}_{-1.99}$ & $4.68^{+6.54}_{-1.72}$ & $10.74^{+0.86}_{-1.33}$ & $5.13^{+16.25}_{-4.23}$ & $28.84^{+22.45}_{-12.99}$ & $11.22^{+12.77}_{-5.05}$& $5308.84^{+61.47}_{-60.77}$ & $8317.64^{+193.74}_{-282.37}$ \\ 
    & & & & & & & & & \\   
    400  & 0.0$^\dagger$ & $4.68^{+0.45}_{-0.60}$ & $7.41^{+1.50}_{-1.10}$ & $10.63^{+0.74}_{-0.87}$ & $1.77^{+17.28}_{-1.53}$ & $8.91^{+13.47}_{-6.16}$ & $6.76^{+2.79}_{-1.27}$& $5308.84^{+61.47}_{-72.84}$ & $8317.64^{+193.74}_{-189.33}$ \\    
    & & & & & & & & &\\
    400  & 1.4$^\dagger$ & $4.68^{+0.57}_{-0.69}$ & $4.68^{+3.26}_{-1.05}$ & $10.38^{+0.89}_{-0.95}$ & $15.85^{+44.41}_{-12.96}$ & $7.08^{+3.68}_{-1.46}$ & $56.23^{+17.89}_{-21.56}$ & $5370.31^{+99.84}_{-98.02}$ & $8317.64^{+193.74}_{-189.33}$\\         
    & & & & & & & & &\\
    400  & 2.0$^\dagger$ & $6.17^{+1.78}_{-1.27}$ & $8.71^{+3.03}_{-4.34}$ & $7.62^{+0.74}_{-0.48}$ & $9.55^{+37.22}_{-8.68}$ & $50.12^{+19.06}_{-11.21}$ & $57.54^{+21.88}_{-22.06}$ & $5321.08^{+74.02}_{-48.78}$ & $8184.65^{+171.38}_{-167.87}$\\
    & & & & & & & & &\\
    \hline
    \end{tabular}}
    \scriptsize{$^\dagger$ Fixed while fitting}\\
    \label{mosfit}
\end{table*}

We modeled the light curves of SN~2019vxm using the interaction model (csm) \footnote{\url{https://mosfit.readthedocs.io/en/latest/index.html}}. The key parameters involved are the mass of the CSM ($\rm M_{CSM}$), ejecta mass ($\rm M_{\text{ejecta}}$), and ejecta velocity ($\rm v_{\text{ejecta}}$). The density profile of the CSM is given by $\rm \rho_{CSM}\propto r^{-s}$, where $s=0$ is a shell profile (eruptive loss) and $s=2$ is a uniform steady wind mass loss. The inner ejecta profile follows $\rm \rho_{ejecta}\propto r^{-n}$, where smaller values of $n$ (\s 7-10) indicate an LBV or WR progenitor, while larger values ($n$\s 12) indicate an RSG progenitor. In the models, the photosphere expands until the minimum temperature $\rm T_{min}$ is achieved, after which it starts to recede. The inner radius of the CSM is given by $\rm R_{0}$ and the CSM density is given by $\rho$. The diffusion time, and thus the rise time, is directly related to the amount of CSM. We use the priors and convergence criteria as given in \cite{2025ApJ...987...13R} and \textit{Dynesty} dynamic nested sampling \citep{2020MNRAS.493.3132S} for the modeling. We have performed modeling for two durations: first, for 200 days as presented in \cite{Lane_2026}, and second, by considering the shallow decline (until t\s 400 days). The later phase (t $>$ 400 days) shows IR brightening that is not accounted for in the model and is therefore not included. We model the UV-optical light curves with a fixed explosion epoch and assuming the efficiency of interaction to be \s 50 $\%$. We also assumed a fully ionized H-rich CSM and fixed the opacity to be $\kappa=0.34\rm\ g\ cm^{-2}$.

We have fixed the parameter $s$, which determines the CSM density profile, and the best-fit values for the different parameters are given in Table \ref{mosfit}. We have fixed $s=0$ and $s=2$ to show two extremes, a shell-like CSM and a wind-like CSM. In addition, we have used $s=1.4$, the best-fit parameter obtained by \cite{Lane_2026}. We obtain CSM mass of $>$4\m\ for all the models with comparable or or slightly higher ejecta masses. We also note the model's dependence on the $s$ values and the extent of the data used for modeling. The CSM mass estimates we obtain also agree with those reported in \cite{2026arXiv260523637L}. However, despite assuming the same $s$ value, the values obtained by us are in stark contrast with those obtained by \cite{Lane_2026}; they obtained $\rm M_{CSM}$ \s 1.48 \m and $\rm M_{ejecta}$ \s 38.8 \m. Further, our models up to 200 days as well as the model by \cite{Lane_2026} do not agree with the data beyond this time, and thus underestimate the parameters. Since we use a short time range that covers the rise but not the decline significantly, the parameters may exhibit degeneracies. Therefore, light curve modeling up to t\s 400 days could provide a better estimate of the parameters. The models also do not account for the IR excess seen in the later phases. Additionally, there could also be some contribution from Ni decay, which is not incorporated in the models. Thus, the mass estimates obtained from the optical light curve, which include only interaction, would have uncertainties due to the other contributors. The model fit for $s=2.0$ until t\s 400 days is shown in Figure \ref{mosfit_lc}. The light curve, along with a comparison of models for different $s$ values, is given in Appendix~\ref{mosfit_apend} along with their posteriors.

\section{Spectroscopic Analysis} \label{spec}

\begin{figure*}[hbt!]
    \centering
     \resizebox{0.8\hsize}{!}{\includegraphics{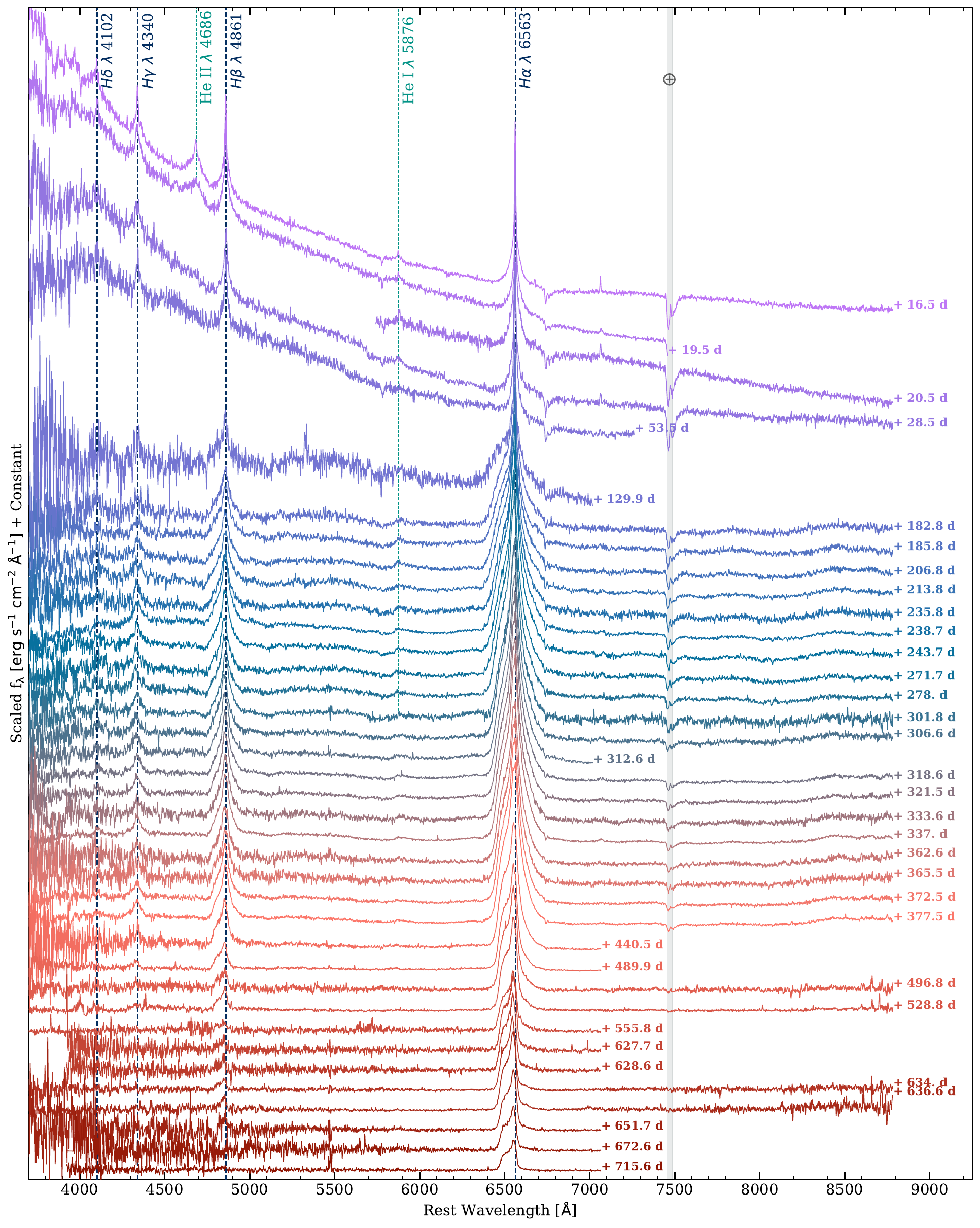}}
    \caption{Spectral evolution of SN 2019vxm from pre-maximum to the late nebular phase spanning about 2 years of observations is shown here. Phases are marked with reference to the explosion epoch (JD$\approx$2458804.5). (All spectra are corrected for redshift, extinction, and absolute flux)}
    \label{spec_evol}
\end{figure*}

Figure \ref{spec_evol} presents the complete optical spectral evolution of SN 2019vxm from t\s 16.5 to 716 days, at 38 epochs. The overall spectral evolution shows a bluer continuum initially (\s 17-54 days) that becomes flatter by \s 130 days. Later on, the continuum is almost flat in the optical until the last epoch of our observations. The spectra are dominated by Balmer lines throughout, with H$\alpha$ being the strongest emission line.

\subsection{Pre-peak and peak spectra}

\begin{figure*}
    \includegraphics[width=0.32\linewidth]{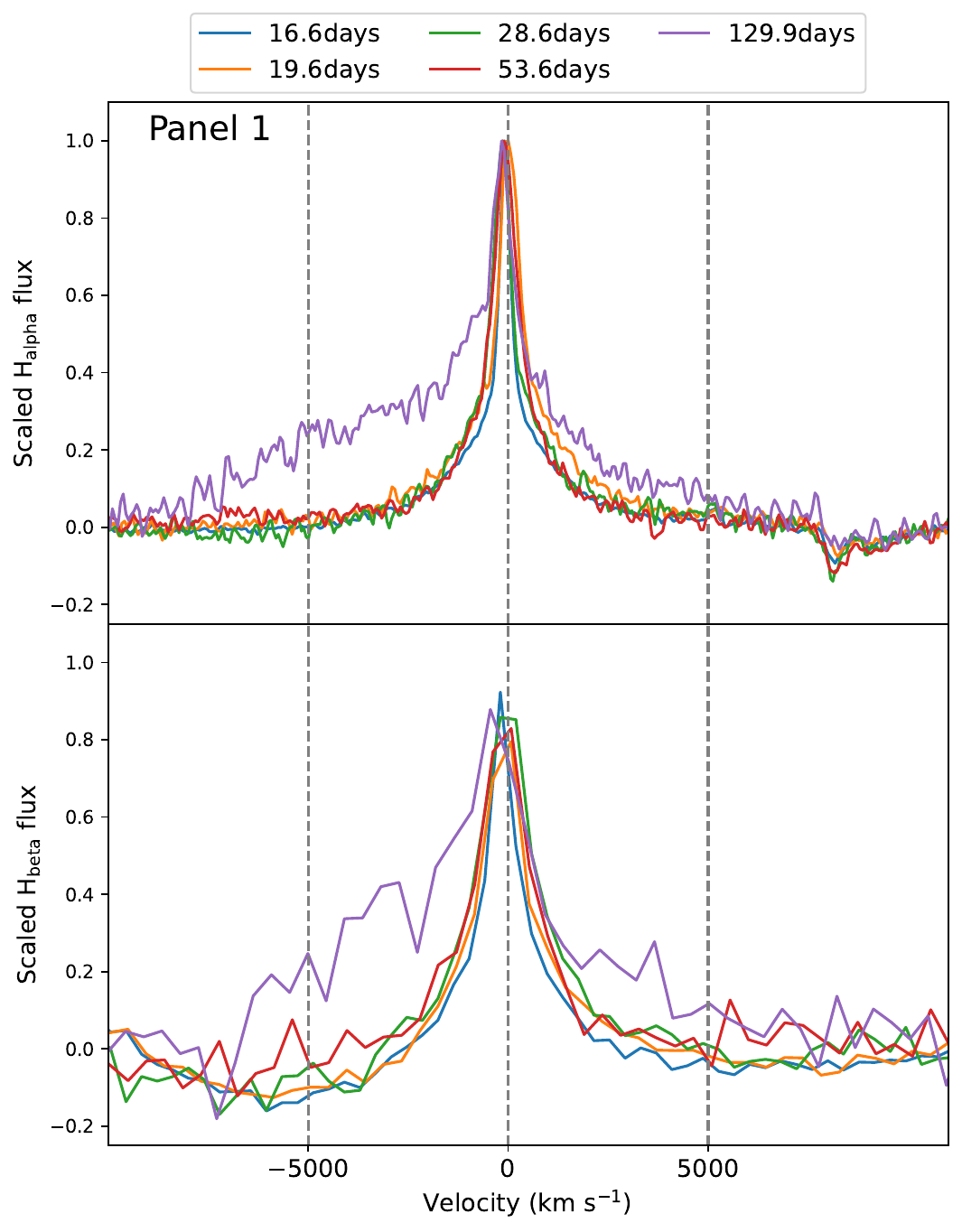}
    \includegraphics[width=0.32\linewidth]{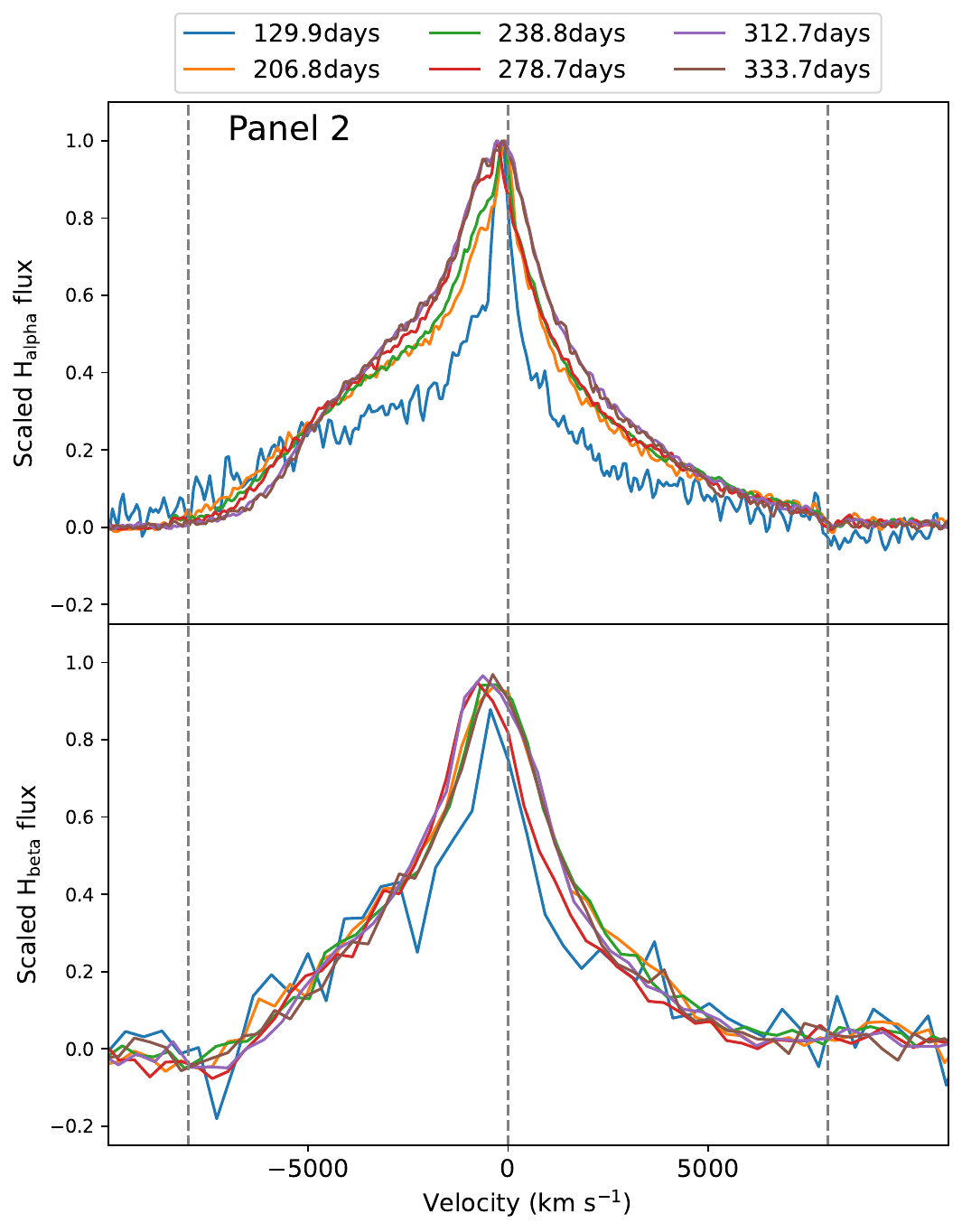}
    \includegraphics[width=0.32\linewidth]{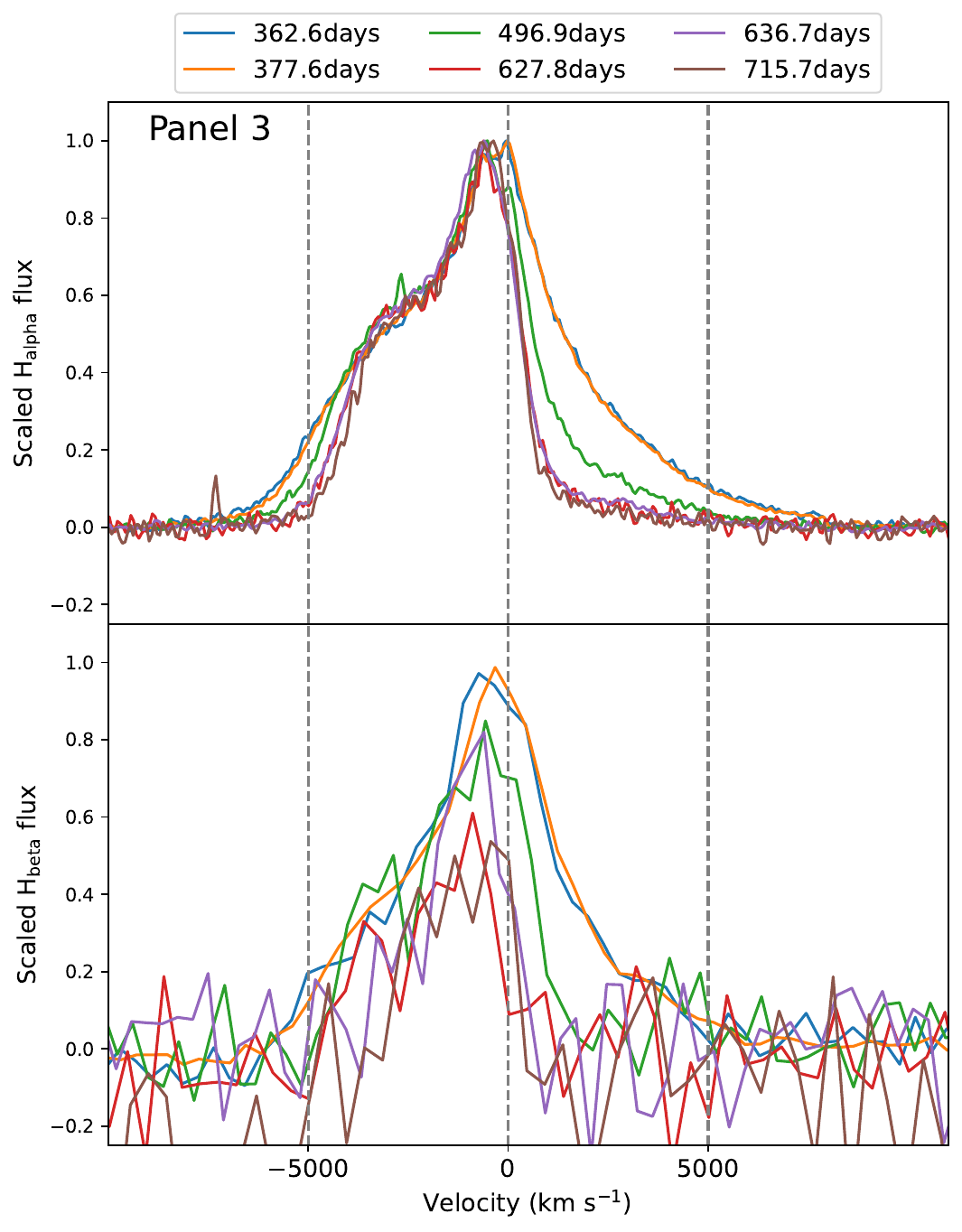}
    \caption{The evolution of H$\alpha$ and H$\beta$. Panel 1: Evolution from t \s 16.5 to 130 days, where initially the profile was symmetric, and the ejecta starts becoming visible at the last epoch. Panel 2: Evolution from t \s 130 days to 334 days. The line asymmetry starts to set in with prominent symmetric wings at \s 8000 \kms (as marked by the vertical lines). Panel 3: Profile evolution from t \s 362 to 715 days, where the clear suppression of red photons can be seen. The wings are still symmetric, but the red wing is more prominent.}
    \label{spec_prof}
\end{figure*}

The first four spectra, from t \s 16.5 days to 28.5 days, represent the pre-peak spectra. They are dominated by narrow emission lines from the shock and intermediate emission lines from the cold dense shell (CDS), which is sweeping up the CSM as it expands. These line profiles are clearly visible in the H$\alpha$ and are also observed in H$\beta$ and H$\gamma$ lines as seen in Figure \ref{spec_evol} and Panel 1 of Figure \ref{spec_prof}. We subtracted the continuum from the profiles by taking an emission-line-free region between $\pm$15,000 and $\pm$10,000 \kms on either side of the center. The H$\beta$ line profiles are less clearly distinguishable due to their relatively lower strength, and the continuum subtraction is also inefficient owing to contamination from neighboring lines. The spectra also show strong signatures of the narrow \ion{He}{1} lines at $\lambda\lambda$ 5876 and 7065 at the first three epochs. The presence of \ion{He}{1} $\lambda$ 6678 is not discernible due to the redder edge of the H$\alpha$ overlapping at those wavelengths. \ion{He}{2}~$\lambda$4686 is also observed in the first two epochs.

\begin{figure}[hbt!]
    \centering
     \resizebox{\hsize}{!}{\includegraphics{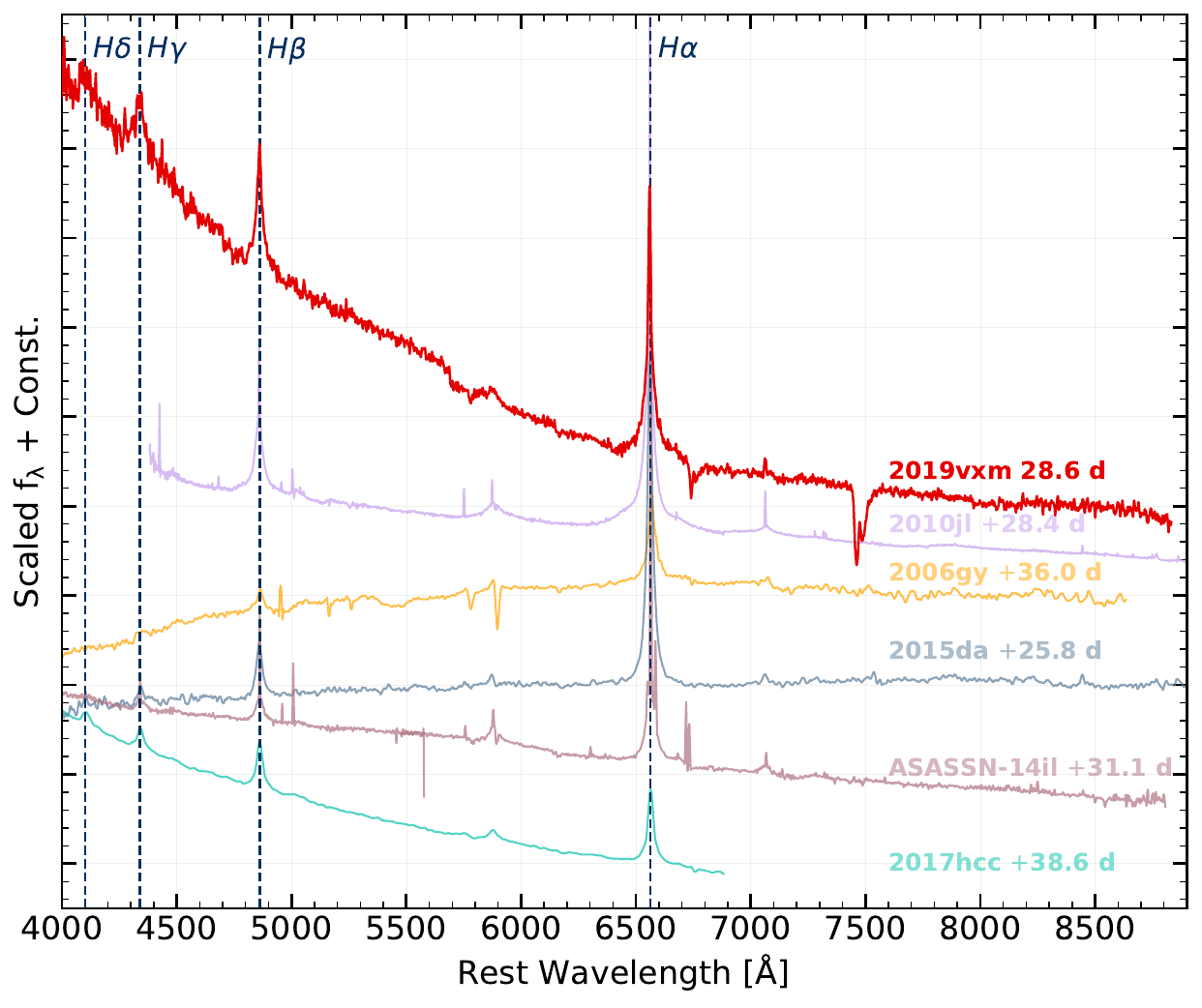}}
    \caption{Spectrum of SN~2019vxm compared with other Type IIn SNe around peak.}
    \label{fig:spec-comp-pre}
\end{figure}

A comparison of the Balmer line profile of SN 2019vxm (refer to Figure \ref{fig:spec-comp-pre}) with those of other IIn SNe at similar phases indicates that such line profiles are very common in this class of objects.  The continuum is also very blue at this epoch, similar to other IIn SNe. The narrow component is also accompanied by prominent wings due to electron scattering, which are fit by a Lorentzian (see Figure \ref{halpha_mirror}). \ion{He}{1} lines are also common in the comparison sample. At these epochs, the hot expanding photosphere is within the CSM. Thus, the dense unshocked CSM in front of the photosphere is the cause for the scattering of narrow line photons \citep{2017hsn..book..403S}.

\subsection{Post-peak spectra}

The spectra obtained between t \s 53 days (maximum) and 372 days (end of the shallow decline) 
also show prominent Balmer lines. The He I and He II lines are no longer observed in the spectra. The spectra taken near maximum show hints of a broad H$\alpha$ component in addition to the narrow and intermediate components. The symmetry in line profiles starts to break down after this epoch (t \s 53 days). The next spectrum was obtained at t\s 130 days, and the broad component is clearly distinguishable, blueshifted by \s 2000 \kms. Panel 2 of Figure \ref{spec_prof} shows the drastic transition from the symmetric to the asymmetric line profile at t\s 130 to 207 days, as evidenced by the H$\alpha$ and H$\beta$ line profiles. The line profile shape changes little over time from t \s 207 to 377 days in this phase. Despite the line profile being asymmetric, the wings, however, are symmetric on either side of the profile (see Appendix~\ref{halpha_max}).

The spectra from t \s 183 days also show weak emission from Ca NIR triplet around 8500 \AA; the emission lines are blended into a single broad feature. Hints of Fe and Si are also seen in the spectra, but are not well distinguished because of the proximity of H$\beta$ and Na ID lines, respectively. Thus, the signatures of the supernova ejecta, if present, are very weak at these epochs except for the strong broad H$\alpha$ emission.

\begin{figure}[hbt!]
    \centering
     \resizebox{\hsize}{!}{\includegraphics{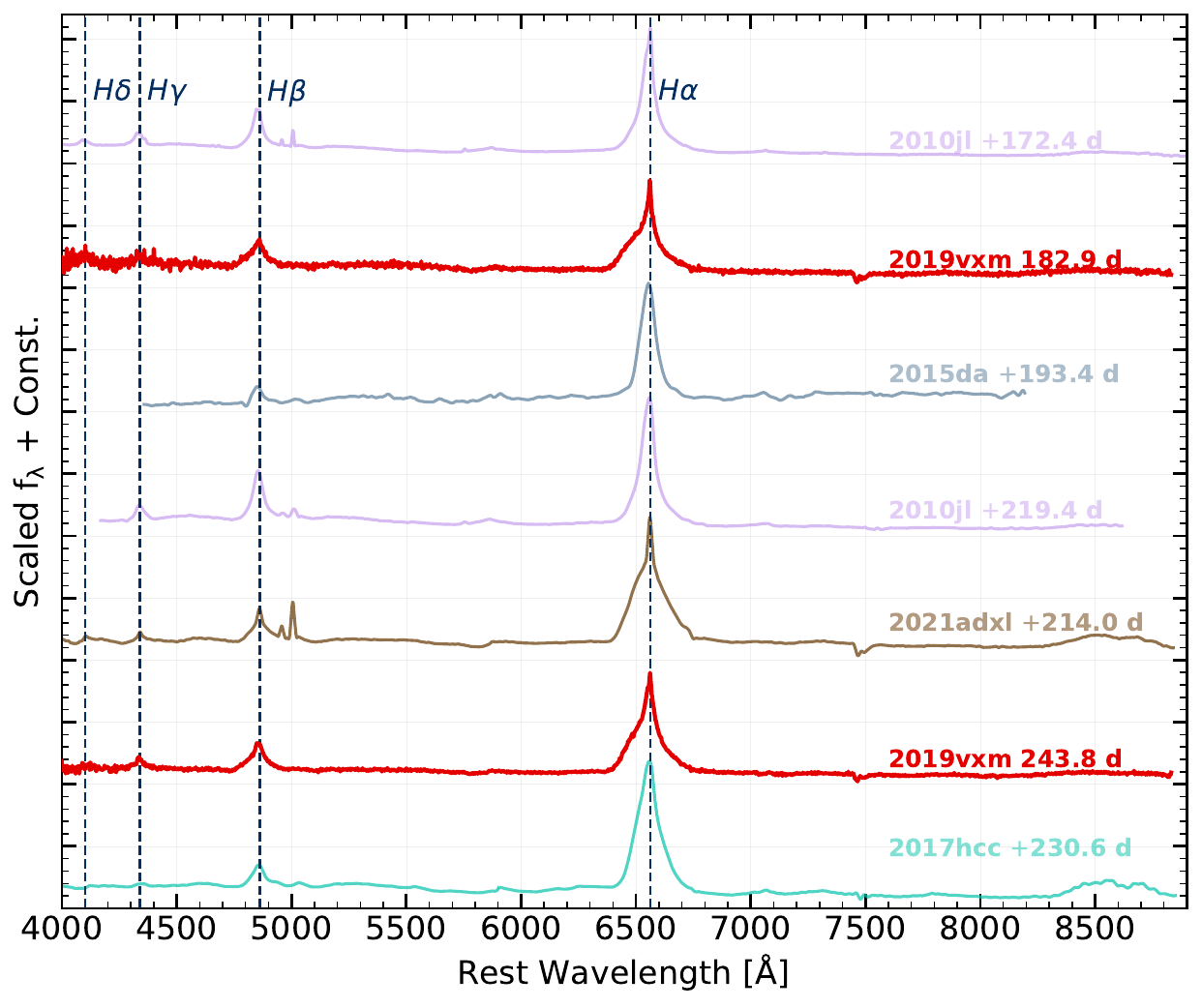}}
    \caption{Spectrum of SN~2019vxm compared with other Type IIn SNe before the linear decline.}
    \label{fig:spec-comp-mid}
\end{figure}

\begin{figure}[hbt!]
    \centering
     \resizebox{\hsize}{!}{\includegraphics{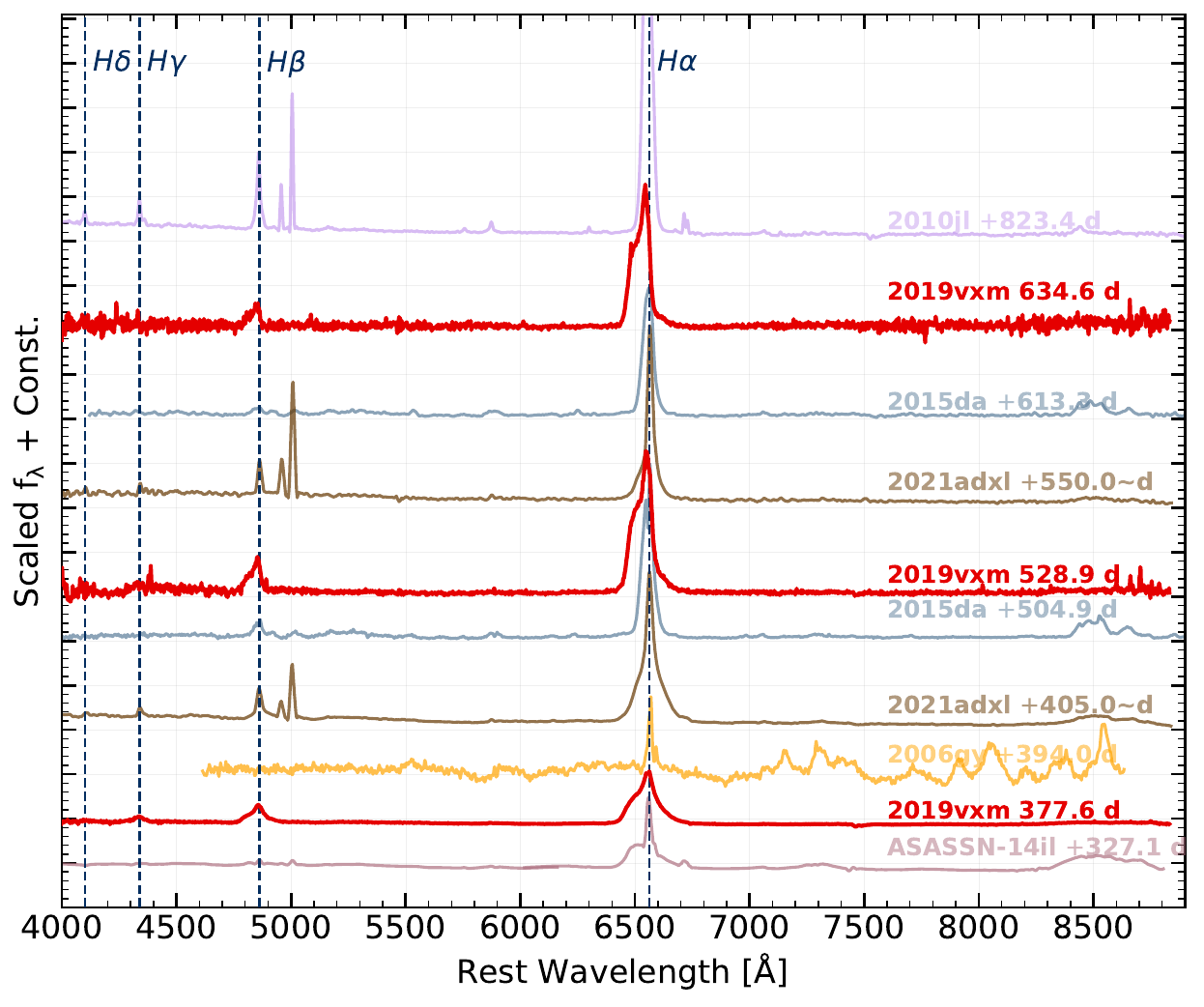}}
    \caption{Spectrum of SN~2019vxm compared with other Type IIn SNe during the late phase decline.}
    \label{fig:spec-comp-late}
\end{figure}

The comparison of SN 2019vxm spectra at t\s 183 and 244 days with the sample of Type IIn SNe in Figure \ref{fig:spec-comp-mid} indicates that, although there is some asymmetry in the line profile, its degree is not uniform. SN 2010jl has a less asymmetric profile, while the H$\alpha$ of SN 2021adxl matches that of SN 2019vxm, with a red wing, a flux deficit at redder wavelengths, and an excess of blue flux, making the line profile appear more skewed. A similar line profile is seen in ASASSN-14il (refer to Figure \ref{fig:spec-comp-late}). The asymmetry of SN 2015da and 2017hcc appears to be between that of SN 2010jl and SN 2021adxl (and therefore SN 2019vxm). The H$\beta$ line profile and its evolution also follow those of H$\alpha$. Weak ejecta features are also visible in SN 2021adxl and SN 2017hcc at this epoch, similar to SN 2019vxm.

\subsection{Late spectra}

The optical spectra from t\s 377 days until t\s 716 lines are dominated by H$\alpha$ and weak H$\beta$ lines. No other prominent line profiles are observed at these epochs; thus, we can conclude that the ejecta remains dense or is significantly obscured. The H$\gamma$ line, which is visible until t\s 500 days, is enshrouded in the noisy spectrum as time progresses. The H$\alpha$ line profile at t\s 377 days is similar to that at the mid-epochs and to those of ASASSN-14il and SN 2021adxl. However, the drastic drop in red flux is only prominent in SN 2015da. This is more conspicuous in the third Panel of Figure \ref{spec_prof}, where the red side flux sharply drops with time. This variation is also seen in the H$\beta$ line profile despite it being weak. However, the H$\alpha$ wings extend to \s 5000 \kms\ on both sides, and are thus symmetric, despite the asymmetric line profile. We discuss more about this blue-red asymmetry in Section \ref{asymm_sec}.

\subsection{Evolution of H$\alpha$ line profile} \label{line_evol_sec}

The H$\alpha$ component was modeled using primarily three components. The first spectrum, taken at t\s 16.5 days, was modeled with a narrow Lorentzian and an intermediate Gaussian, indicating emission from the CSM and the CDS.  The narrow component corresponding to the CSM is not resolvable, and thus, we could only constrain the upper limit of the velocity. As the SN rises to its maximum, from t \s 40 days a broad ejecta component of $\rm v \sim 13,000\ km\ s^{-1}$ starts becoming visible, and is prominent in the next spectrum taken \s 76 days later. For t \s 40 days, we do not require a Lorentzian profile to model the narrow component; a Gaussian profile suffices, since the wings are less prominent and are partly already included in the broad component. After t$>$130 days, we also observe a red tail and a blue excess, which are initially modeled by three components. When the narrow line is not detectable anymore from t \s 271 days, we use two Gaussian profiles for the intermediate and broad components.

In the initial phases, the H$\alpha$ was dominated by emission from the CDS and the CSM, while the ejecta flux became noticeable near peak. The ejecta flux contribution increases, reaching a maximum at t\s 239 days and then falls from t\s 337 days, thus showing prominence near the shallow decline phase. The narrow component always has an FWHM below the resolution limit; therefore, we take 600 \kms as the upper limit of the velocity. The CDS, initially moving at a velocity of v \s 4000 \kms, decelerates as the ejecta becomes visible, and evolves with a velocity of v \s 2000 \kms. The ejecta, initially fast moving \s 15000 \kms\ at t\s 50 days, decelerates rapidly, reaching a velocity of \s 8000 \kms\ at t\s 200 days, after which it decelerates very slowly to \s 7000 \kms\ by t \s 300 days. This is in agreement with the ejecta velocity of \s 8300 obtained from MOSFiT models (refer to Table~\ref{mosfit}). We did not perform further modeling beyond t \s 400 days due to pronounced asymmetry that suppresses the redder flux and leads to miscalculations. We also stress that the velocities obtained from the fit are subject to uncertainties, as we have not accounted for the blue excess or the red deficit.

From the prominent broad H$\alpha$ line flux, an oversimplified estimate of the ejecta mass can be obtained. We used the flux at t \s 270 days to obtain the electron density using:
\begin{equation}
    f=\epsilon n_{e}n_{i}Vd^{2}
\end{equation}
where f is the flux, $\epsilon$ is the emissivity, $n_e$ and $n_{i}$ are the electron and ion densities, $V$ ($V=(4/3)\pi (v_{exp}t)^{3}f_{r}$) is the volume, and $d$ is the distance to the SN. Here, we use the volume corresponding to the extent to which the ejecta has expanded at this epoch ($v_{exp}t$), and we assume a volume-filling factor ($f_{r}$) of 0.1. The recombination coefficient was obtained from \cite{1987MNRAS.224..801H} assuming T=5000 K for calculating the emissivity. The mass of the ejecta ($M_{ej}$) was further calculated using:
\begin{equation}
    M_{ej}=n_{e}m_{H}V
\end{equation}
where $m_{H}$ is the mass of hydrogen. Thus, we estimate a lower limit of 3.88 \m\ for the ejecta mass. This is lower than the ejecta mass estimates for Type II SNe (5-30 \m) \citep{2024Ap&SS.369...49U} and comparable to those obtained for Type IIn SNe by \cite{2026A&A...706A.169E} through population synthesis models. For low luminosity Type IIP SNe, the ejecta mass range is from 5.7-12.9 \m\ \citep{2026PASP..138b4204D}. Short-plateau IIP SNe have ejecta mass varying from a few solar masses to up to upto 25 \m \citep{2021ApJ...913...55H,2022ApJ...930...34T,2024Ap&SS.369...49U}. A substantial amount of stripping in SNe IIn can essentially support a low ejecta mass inferred from the H$\alpha$ line.

\subsection{Asymmetry} \label{asymm_sec}

\begin{figure*}
    \begin{minipage}[t]{0.55\textwidth}
        \includegraphics[width=0.5\linewidth]{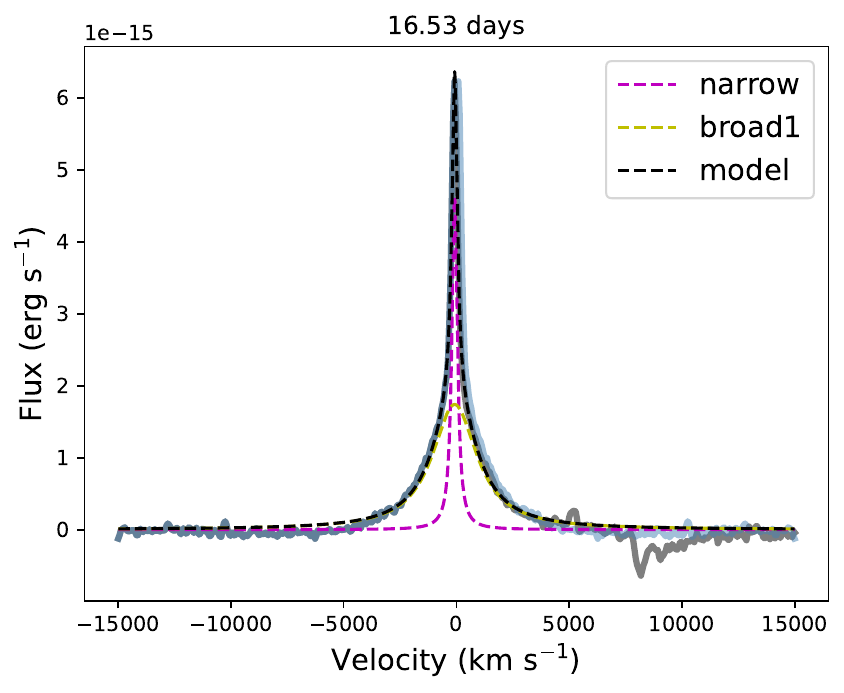}
        \includegraphics[width=0.5\linewidth]{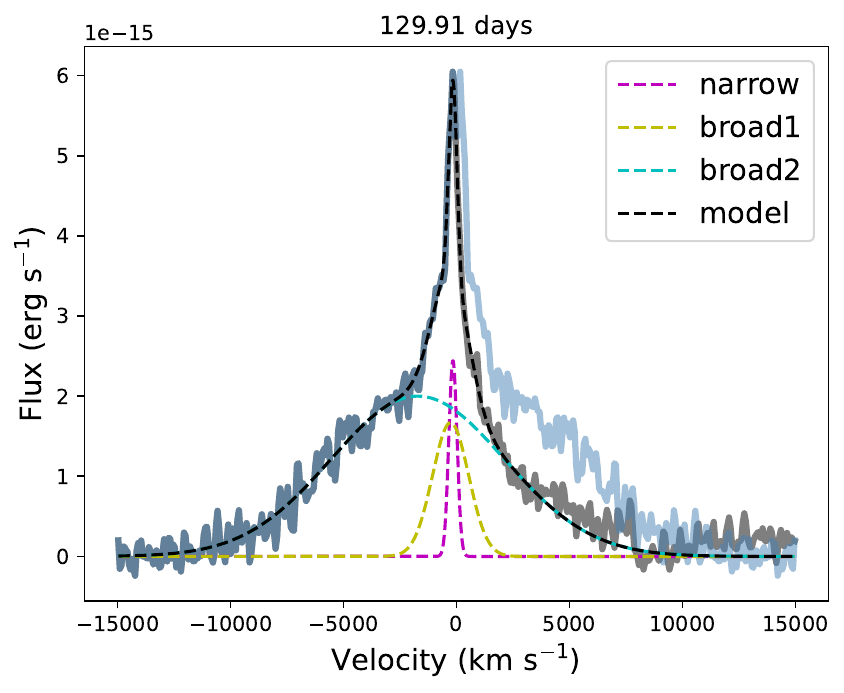}
                
        \includegraphics[width=0.5\linewidth]{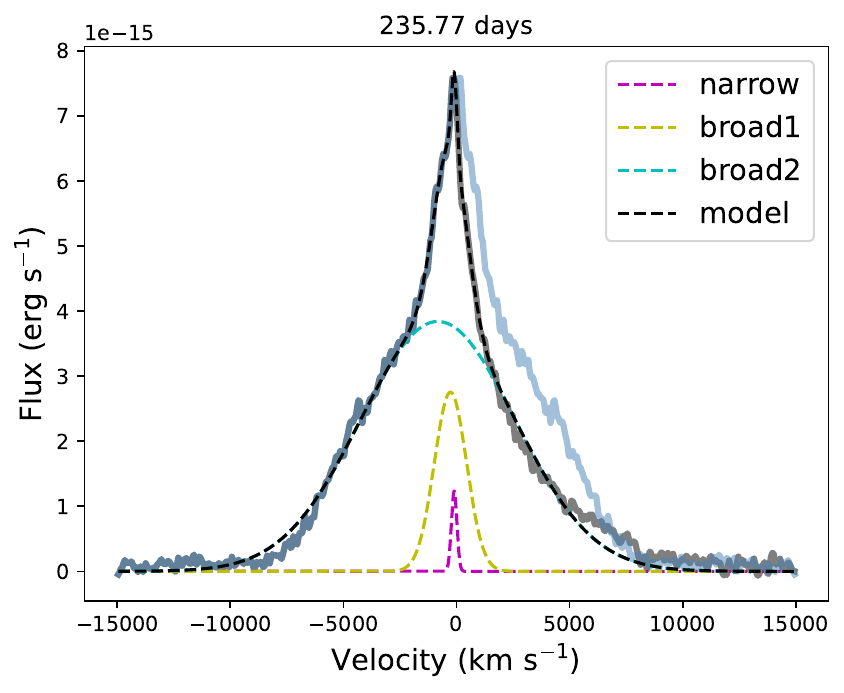}
        \includegraphics[width=0.5\linewidth]{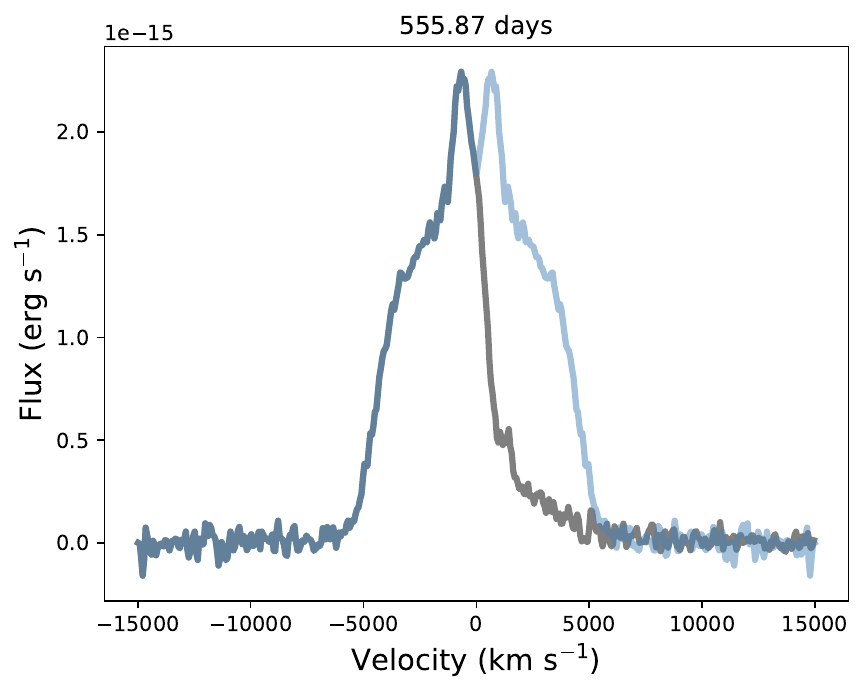}
    \end{minipage}
    \hfill
    \begin{minipage}[t]{0.85\textwidth}
    \vspace{-3.5cm}
    \includegraphics[width=0.55\linewidth]{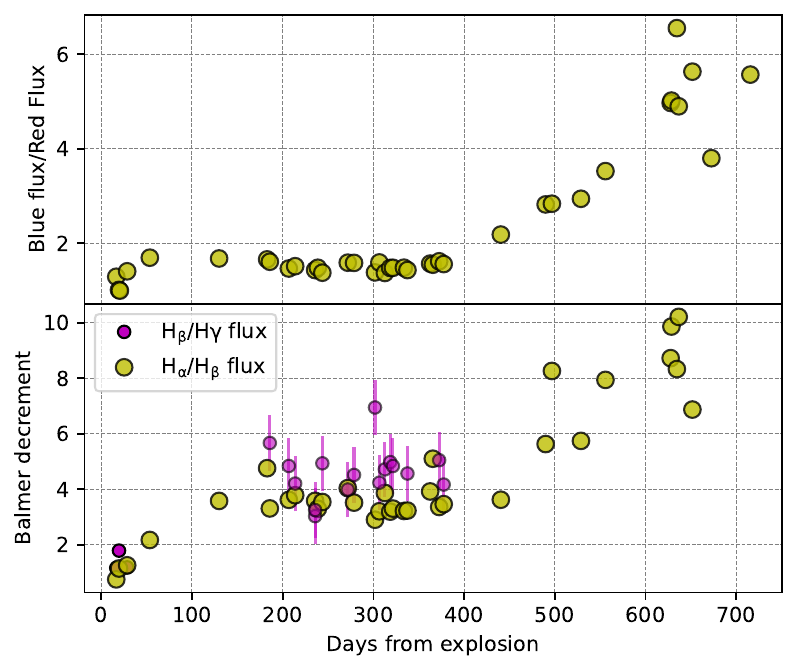}
    \end{minipage}
    \caption{\textbf{Left: }Evolution of H$\alpha$ line profile as time evolves. The grey spectrum is the observed one. The flux at the blueshifted velocities is mirrored at the redder end to indicate the degree of flux excess and deficit in blue. The later epochs also indicate a prominent blueshift of the emission line, as the blue excess, when mirrored, appears as two peaks. The Lorentzian and Gaussian line profiles used for modeling are also included in the Figure. \textbf{Right: }The evolution of asymmetry calculated as the flux in blue side to red side (top) and the Balmer decrement (bottom).}
    \label{halpha_mirror}
\end{figure*}

The presence of line asymmetry is not uncommon in SNe IIn, as previously discussed in Section \ref{spec}. However, to qualitatively fathom the degree of asymmetry, we can compare the flux on either side. To represent the flux deficit, we have mirrored the blue side on both sides of the zero-velocity, as shown by the blue profile in Figure \ref{halpha_mirror}. As is evident from the evolution, the line profile is symmetric in the initial phases (t \s 16.5 days). However, it becomes asymmetric, and a red deficit is observed from t \s 130 days. The degree of asymmetry is visually the same from t \s 200 and 400 days (similar to the profile in Figure \ref{halpha_mirror} for t \s 238 days). There is a red deficit, and the red wing is more pronounced than the blue. But the spectra taken from t \s 400 days show a more obvious asymmetry, with a sharper deficit of red photons. The red wing seen in earlier epochs damps down, and the blue side appears more boxy. The line profile is also blueshifted relative to the previous epochs. The blue-red asymmetry of H$\alpha$ in the spectral line profile is also quantitatively shown as the ratio of the flux in the blue side to the flux in the red side in the right panel of Figure \ref{halpha_mirror}. The ratio gradually increases until t\s 40 days, then remains constant until t\s 400 days. It then rises rapidly from t \s 426, which corresponds to the same epoch where the optical light curve declines sharply, and the IR excess starts appearing. We discuss this in detail in Section \ref{dust}.

\section{Discussion} \label{discuss_sec}

\subsection{The SN environment: the properties of the CSM}

The bolometric light curve of the SN is initially dominated by UV reaching a peak luminosity $\sim\rm 5.23\times 10^{43}\ erg\ s^{-1}$, which is comparable to $\sim 3\times 10^{43} \rm\ erg\ s^{-1}$ of SN 2010jl \citep{2014ApJ...797..118F}, even though its evolution follows a completely different trend. This is \s 10$\%$ of the bolometric luminosity calculated for SLSNe like SN 2016aps \citep{2021ApJ...908...99S} and \s 27 $\%$ of the bolometric luminosity calculated for SN 2015da \citep{2024MNRAS.530..405S}. \cite{2024arXiv241107287H} investigated the radiated energy distribution and found a clear bimodality centered near \s $10^{49}$ erg and \s $2 \times 10^{50}$ erg. So SN 2019vxm falls in the highly energetic category among IIns. In the late phases, the optical light curve declines at a rate similar to radioactive Ni-Co decay, assuming complete $\gamma$-ray trapping. Modeling the entire duration of multiband light curves using the \textit{CSM+Ni} decay model in MOSFiT has proved unfruitful. Including the IR contribution, suggests that the mechanism powering the SN in the late phases is not Ni decay, also obtained by \cite{2026arXiv260523637L} through modeling. This dominance of IR luminosity is seen in multiple IIn SNe, including SN 2010jl, where the IR rises from 33$\%$ to 50$\%$ to 85$\%$ at t\s 100, 400, and 800 days, respectively. 

\begin{figure}[b]
    \centering
    \includegraphics[width=1\linewidth]{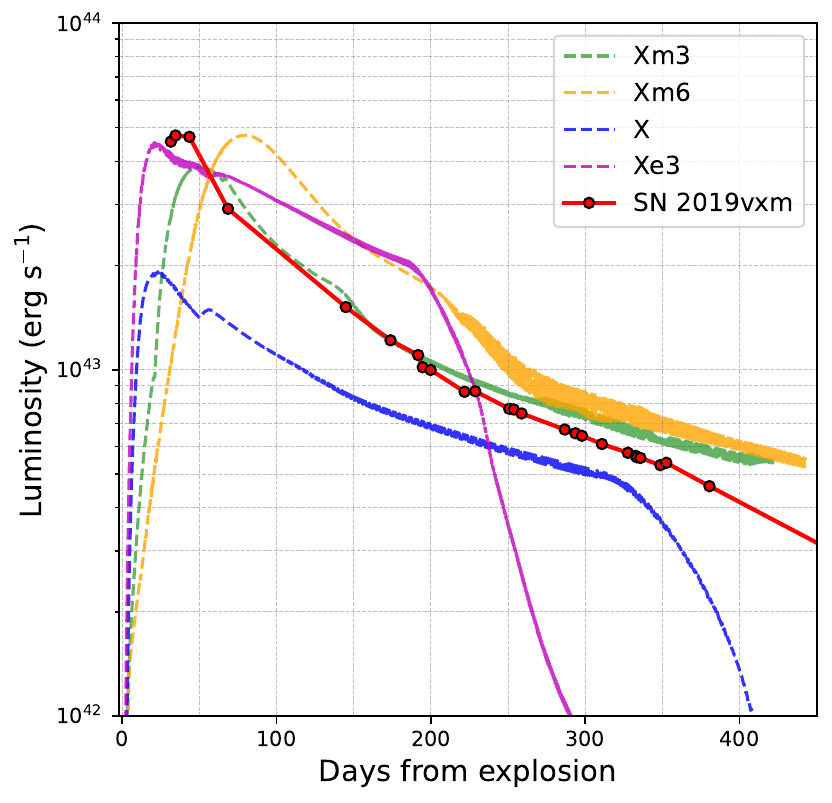}
    \caption{Comparison of SN 2019vxm (red) with results of radiative transfer models given by \cite{2015MNRAS.449.4304D}, details of which are mentioned in the text.}
    \label{dessart_mod}
\end{figure}

The CSM mass estimated for the interaction of SN 2010jl is $\gtrsim$3 \m \citep{2014ApJ...797..118F}, while \cite{Lane_2026} estimated a mass of $\rm M_{CSM}$ \s 1.5 \m for SN 2019vxm from $s=1.4$. However, the model is unable to explain the data at later phases. We obtain higher masses for our models, but the fits are poor, with large errors, as shown in Appendix~\ref{mosfit_apend}. Despite convergence, the model degeneracies arising from the large number of free parameters make such estimates highly dependent on the data and the prior. Additionally, constraining $s$ would indirectly place bounds on the CSM mass, ejecta mass, density, and $\rm R_{0}$. Moreover, IR brightening is prominent at late phases; thus, the estimates we obtain with MOSFiT are lower. Therefore, we compare the bolometric luminosity we obtained with the theoretical bolometric luminosity evolution presented by \cite{2015MNRAS.449.4304D} \footnote{\url{https://doi.org/ 10.5281/zenodo.5524978}}. The authors used a radiative transfer model implemented in HERACLES to obtain the light curve based on properties of the well-studied SN 2010jl. The model considers various CSM and energy density combinations; they also have more luminous models, but since we are using it for comparison with SN 2019vxm, those are not included. Figure \ref{dessart_mod} represents the evolution of bolometric luminosities for four models, X, Xe3, Xm3, and  Xm6. X corresponds to the model with kinetic energy $10^{51}$ erg while Xe3 has an energy of $3\times 10^{51}$ erg; the CSM mass and mass loss rates are 2.89 \m\ and 0.1 \mloss\ for both. Xm3 and Xm6 have the same kinetic energy as X, but the mass loss rates are $\rm \dot{M}\sim 0.3$ and 0.6 \mloss, respectively. These mass loss rates are for the inner CSM (R$<10^{16}$ cm), while the mass loss rates are two orders lower at R$> 10^{16}$ cm. We see that the peak luminosity matches that of Xe3 and Xm6, with a slightly longer rise time than Xe3. Considering the dynamics at play, where the bolometric luminosity depends on both the intrinsic KE and the energy tapped during interaction (related to the CSM mass, keeping the efficiency constant), we can say that this SN is less energetic than Xe3 but with a more massive CSM. This is inferred from the longer rise time of SN 2019vxm compared to Xe3, indicating a more massive CSM. However, the rise time is slightly less than that of Xm3. The total radiated energy calculated from the HERACLES simulation for Xm3 is $\sim\ 4.9\times 10^{50}$ erg, which agrees with the obtained value for SN 2019vxm. Thus, the CSM mass can be between \s 2.8-8 \m. In the comparison at later phases, where X and Xe3 models decline steeply, the bolometric luminosity of SN 2019vxm shows a uniform decline. This strengthens the argument for a massive CSM that extends farther out. 

We can also obtain an approximate mass loss rate using:
\begin{equation}
    \dot{M}=\frac{2L}{w}\frac{v_{CSM}}{v_{sh}^{3}}
\end{equation}
at each epoch \citep{1994MNRAS.268..173C,2024MNRAS.530..405S}, with L denoting the luminosity, $\rm v_{CSM}$ is the CSM velocity set to 100 \kms, w represents the efficiency of conversion of KE to luminosity, and the forward shock velocity $\rm v_{sh}$ is determined from the initial velocity of the intermediate component and taken as 4000 \kms. Integrating it over time, $\rm t=t_{SN}\times v_{sh}/v_{CSM}$ up to $T_{SN}$ \s 612 days (t $_{SN}$ is the time taken from the shock to traverse to that region) gives an estimated mass of \s 5.91 \m\ assuming 50$\%$ efficiency. The same calculation for SN 2015da yielded a mass of \s 20 \m\ and 5.6 \m\ for ASASSN-14il \citep{2024MNRAS.530..405S,Dukiya_2024}.

Thus, the CSM mass from MOSFit, the analytical model, and the luminosity-mass relation suggest that SN 2019vxm has a massive CSM. However, the ejecta masses are relatively low. These could be due to intrinsically low ejecta or to modeling degeneracies. Since the model involves multiple parameters, some of which are dependent on one another, constraining $s$ would indirectly place bounds on the CSM mass, ejecta mass, density, and $\rm R_{0}$. A smaller $s$ indicates a denser CSM; they are inversely related to each other. The ejecta mass and CSM mass also follow an inverse relation; assuming the same conversion efficiency and ejecta velocity, a more massive ejecta, owing to larger kinetic energies, would require less CSM than less massive ejecta to produce the same energy output. Additionally, if the efficiency of KE conversion is lower than what we have assumed, the KE associated with it, and thus the ejecta mass, could be larger. From the broad H$\alpha$ component, we calculated the lower limit of ejecta mass as \s 3.88 \m\ (Section~\ref{line_evol_sec}). This is in agreement with all the models and with \cite{Lane_2026}. However, \cite{2026arXiv260523637L} obtains an ejecta mass of 26-100 \m.

The deconvolution of the H$\alpha$ line profile up to t\s 400 days indicates an enhanced flux solely due to interaction in the initial phases; the flux from the intermediate and broad components dominates near the maximum. The broad component starts becoming apparent at t \s 50 days (though the significance is very low) and starts becoming stronger by t \s 130, reaching a maximum flux of $\rm \sim 1.2\times 10^{-11}\ erg\ s^{-1}$ at t \s 250 days, after which it fairly stays the same and then declines slowly from 300 days. The contribution of the intermediate flux does not change prominently during the evolution and is $\rm \sim 2\times 10^{-12} \ erg\ s^{-1}$. We further note that the phase when the flux from the broad component becomes dominant also corresponds to the shallow decline in the light curve. The spectra also show signatures of ejecta, with weak Ca NIR triplets at $\lambda\lambda\lambda$ 8498, 8542, and 8662 \AA, and possibly other lines that could be blended with the strong Balmer features at the same epoch. The narrow component is not resolvable from t \s 250 days, indicating that either the unshocked CSM is not sufficient to produce strong narrow lines or that they are unresolvable due to our instrument limitations. The broad component decelerates slowly, attaining a constant value, which is also followed by the intermediate component. We do not attempt to make quantitative comparisons of the velocity, taking into account the uncertainties due to asymmetry.

\subsection{Pre-explosive mass loss and detection}

The binned ATLAS light curve shows no enhanced mass-loss episodes over the 4 years preceding the explosion. \cite{Lane_2026} also observed no detected precursor in ATLAS, TESS, and Pan-STARRS observations. However, we can estimate the expected magnitude from the CSM mass obtained and infer whether such mass loss and timescales can indeed lead to a detectable precursor.

The extent of CSM traced by t \s 612 days is $2.1 \times 10^{16}$ cm assuming a CDS velocity of 4000 \kms throughout, which, on assuming a wind velocity of 100 \kms, corresponds to mass lost \s 67 years before explosion. Considering a uniform mass loss, the average mass-loss rate is \s 0.088 \mloss. We present two scenarios for enhanced luminosity corresponding to such mass losses: an interaction-powered luminosity or a line-driven wind \citep{2014ApJ...789..104O,2021ApJ...907...99S}.

If we assume a line-driven wind in a super-Eddington system leading to instability in the atmosphere given by \cite{2001MNRAS.326..126S}, we can relate the mass loss rate and luminosity as:
\begin{equation}
    L_{rad}=\frac{\dot{M}_{CSM}c_{s}c}{W}
\end{equation}
where $L_{rad}$ is the radiated luminosity, $\dot{M}_{CSM}$ is the mass loss rate, $c_{s}$ is the speed of sound taken as 60 \kms, c is the speed of light and W$\approx$ 5 is a constant. The radiated energy then is $2.1 \times 10^{41}\rm erg\ s^{-1}$.

On the other hand, for an interaction-powered precursor, the kinetic energy would be converted to radiation as:
\begin{equation}
    L_{rad}=\epsilon\frac{1}{2}\dot{M}_{CSM}{v_{CSM}}^{2}
\end{equation}
where $\epsilon$ is the efficiency of conversion and $v_{CSM}$ is the CSM velocity during the mass loss. A population study of precursors by \cite{2021ApJ...907...99S} has shown that, by comparing the luminosity of the explosion and the precursor, the estimate of $\rm v_{CSM}$ during the mass loss is \s 2000 \kms. Then the luminosity, assuming $\epsilon=1$, would be \s $1.17\times 10^{41}\rm\ erg\ s^{-1}$. In this specific case, the $v_{CSM}$ estimate is still unknown; however, since the luminosity is lower than in the former case, we can use that to obtain magnitude estimates.

Since the nature of the progenitor is unknown, we have used a range of temperatures from 3000 K to 10000 K for the progenitor. We assume the CSM has approximately the same temperature as the star. Thus, using the luminosity obtained, one can determine the flux density assuming a blackbody.  The magnitudes in specific filters were obtained using \textit{sncosmo}, incorporating the filter response of the instrument. Thus, we obtain ATLAS-\textit{o} band magnitude, SDSS-\textit{r} magnitude, and ZTF-\textit{r} band magnitude corresponding to such mass loss to be 22.12, 22.6, and 22.42~mag for 3000 K and 21.88, 21.82, and 21.84~mag for 10000 K, respectively. This can explain why we did not detect any precursors for SN 2019vxm. If the mass-loss rate were $>$ 0.24 \mloss, we would obtain a magnitude $<$ 20 for the precursor (the luminosity needed is \s 5.78$\times 10^{41}$ erg), making it detectable in the wind-driven mass-loss scenario. On the other hand, for the interaction-driven precursor, we would need $\dot{M}$ \s 0.88 \mloss\ assuming $\epsilon=0.5$. However, this depends on the velocity and efficiency.

\subsection{Scattering dominated evolution}

Prominent line asymmetry sets in by t \s 130 days as seen in Figure \ref{halpha_mirror}. The first figure (left) shows fairly symmetric line profiles, while the second (center) shows blueshifted emission with large velocities. This can be explained by the occultation of red photons by the receding photosphere \citep{1976ApJ...207..872C}; in that case, the prolonged asymmetry should decrease over time. The SED evolution at this epoch also corresponds to the photosphere beginning to recede into the ejecta from the unshocked CSM. The profile from t \s 206 to 377 days (see Figure \ref{spec_prof}), however, shows asymmetry where the H$\alpha$ profile does not change with time. We also note prominent wings in the redder part; they are of equal extent on either side, but the profile is skewed. Appendix ~\ref{halpha_max} presents a table of the maximum velocities measured from the H$\alpha$ line profile, showing that the extent on both sides is symmetric about the rest wavelength. In case of SN 2019vxm, we do not observe any blueshift in the peak emission line profile, in agreement with \cite{2001MNRAS.326.1448C}, who concludes that the blueshift is very small (\s 30 \kms) in such electron scattering scenarios, but this could also be due to resolution limitations. 

The line asymmetry observed until t \s 380 days is not unique to SN 2019vxm. The profile is similar to that of SN 2015da, and there are two explanations put forward for the asymmetry: \cite{2020AA...635A..39T} adopts an electron scattering model, while \cite{2017hsn..book..403S} claims occultation by the CDS. The H$\alpha$ profile is very similar to that of SN 2021adxl, for which \cite{2024AA...690A.259B} successfully modeled the emission by assuming the intrinsic emission arises from the CDS and undergoes scattering as it propagates through the environment. Such line profiles are also observed in SN 2010jl, which \cite{2014ApJ...797..118F} claim to be due to electron scattering, while \cite{2020MNRAS.499.3544S} claim the same line profile due to obscuration by CDS. However, all these scenarios are equally possible and will be discussed further in Section \ref{dust}.

Flux ratios, including those of Balmer lines like $H_{\alpha}/H_{\beta},\ H_{\beta}/H_{\gamma}$ and Paschen lines, are indicators of the properties of the line-emitting region \citep{1989agna.book.....O}. Figure \ref{halpha_mirror} represents the evolution of $\rm H_{\alpha}/H_{\beta}$ and $\rm H_{\beta}/H_{\gamma}$ flux with time. Here, since both H$\alpha$ and H$\beta$ are strong, we use the flux under the line to calculate the ratio, while for H$\gamma$, we scale the profile to that of H$\beta$ to estimate the values due to low SNR. Since it is manually done for the latter, we use a scaling step of 0.5; the error included here is the scaling step. The Balmer decrement rises and then settles into a constant value of \s 3-4 from t \s 130 to 400 days. Such decrement values are also observed in SN 2010jl and may be attributed to scattering by a dense or optically thick medium, assuming Case B recombination \citep{1989agna.book.....O,1980ApJS...42..351D}. The constant Balmer decrement indicates that the properties of the obscuring scattering medium remain unchanged during that phase. The Balmer decrement values matches with the theoretical predictions for electron temperature $T_{e}=(1.2-1.6)\times 10^{4}$ K and electron number density $N_{e}=10^{9}-10^{10}\rm\ cm^{-3}$ \citep{1980ApJS...42..351D} $T_{e}$. It also is in agreement with values calculated for $T_{e}=1000$ K and $N_{e}=10^{2}-10^{4}\rm\ cm^{-3}$ \citep{1987MNRAS.224..801H}.
Prior studies show that the Balmer decrement of SNe like SN 2010jl, ASASSN-14il, and SN 2015da increased steadily \citep{Dukiya_2024,2014ApJ...797..118F}), in contrast with SN 2019vxm, where the Balmer decrement shows little or no variation from t \s 150 to 400 days, after which it steadily increases.
The sudden increase in Balmer decrement as well as asymmetry ratio indicates that there is a change in properties of the extincting medium, and the change is ongoing until the last epoch of observation at t \s 715 days.

\subsection{Dust}  \label{dust}
\begin{figure}
    \centering
    \includegraphics[width=1.0\linewidth]{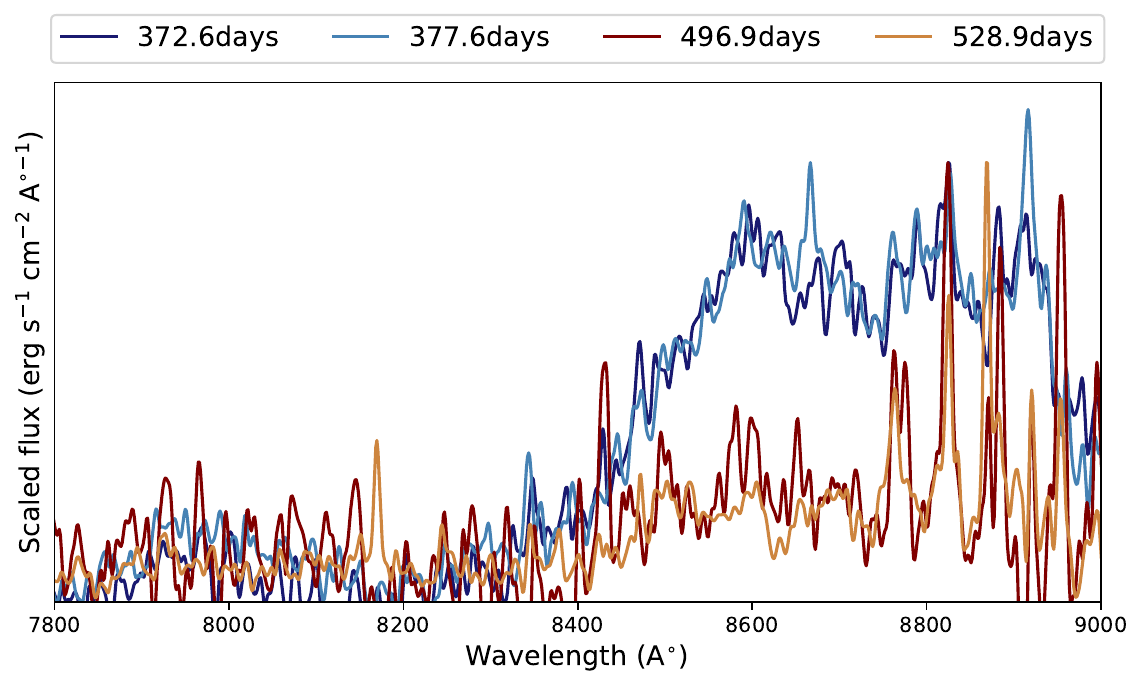}
    \caption{The evolution of Ca NIR triplets at t\s 373, 378, 497 and 529 days.}
    \label{ca_lines}
\end{figure}
The IR excess seen in the SED (Figure \ref{SED}), coupled with a marked asymmetry in line profiles and drop in optical LC, is a telltale signature of the influence of dust, either pre-existing or newly formed. Unlike SN 2017hcc, where \cite{2023AA...669A..51M} is unable to detect the transitioning phase due to a lack of observations during solar conjunction, we can observe the smooth evolution of the line profiles. The line profile narrows, indicating deceleration of the ejecta, but it also becomes more boxy. In addition to the steady rise in asymmetry, we also observe a steady increase in the Balmer decrement. But since we only have H$\alpha$/H$\beta$, we cannot infer the properties of the extinction material, where we would expect wavelength-dependent extinction. However, there is a clear blueshift in the line profiles before and after the IR becomes dominant, strengthening the claim of dust being present. The IR light curves also show a decrease in magnitude in the \textit{H} and \textit{Ks}-bands at t \s 871 days. The \textit{H-K} color at this epoch is 1.44, which supports both the pre-existing and newly formed dust cases. Assuming the dust to be composed of C grains of $\rho \sim 2.25\rm \ g\ cm^{-3}$ and $<1\mu$m size, we can estimate the mass as \s (0.85-1.42) $\times 10^{-4}$ \m \citep{2017MNRAS.466.4221E}.

Pre-existing dust in the CSM can be heated by the expanding shock, provided it has not evaporated during the explosion. \cite{2014ApJ...797..118F} calculated the evaporation radius for SN 2010jl using the maximum luminosity of $\rm L_{max}\sim 3\times 10^{43}\ erg\ s^{-1}$, which is comparable to SN 2019vxm. The evaporation radius obtained for SN 2010jl is $\rm R_{evap}\sim 3.4\times 10^{16}$ cm and $\sim 7.2\times 10^{16}$ cm for $\rm T_{eff}$ \s 6000K and dust grain size of 1 $\mu$m for graphite and silicate dusts, respectively \citep{1979ApJ...231..438D}. For SN 2019vxm, the values would be 1.5 times higher due to its higher luminosity. From SED modeling, we get the radius of the NIR emitting component to be $\sim 4.0\times 10^{16}$ cm. Thus, the presence of pre-existing dust is entirely plausible. \cite{2026arXiv260523637L} has used the maximum velocity from the photospheric radius to show that the shock can reach this radius and thus contribute to the IR excess and the pronounced asymmetry at this phase. They also observe an MIR excess at t$<$ 400 days, which can be attributed to IR echo, not prominent in the JHK bands as presented. In that case, the constant Balmer decrement could be due to scattering by low-temperature dust; however, we cannot conclude whether the extinction is wavelength-dependent due to poor H$\gamma$ SNR. Evidence of pre-existing dust is present in many Type IIn SNe that have been extensively followed up in the IR \citep{2011ApJ...741....7F}.

Newly formed dust can also account for most of the optical and spectral data obtained for SN 2019vxm. The increase in the Balmer decrement and asymmetry could be due to newly formed dust, which preferentially absorbs at shorter wavelengths. This can also explain the Ca features visible at t \s 377 days but not at t \s 496 days (refer to Figure \ref{ca_lines}), since dust obscuration would mask the ejecta. Thus, in this case, the early phase is dominated by scattering by an optically thick dense medium, while the latter phase is dominated by IR emission from newly formed dust. The H$\alpha$/H$\beta$ and H$\beta$/H$\gamma$ ratios being equal could possibly hint at wavelength-independent scattering. However, the poor SNR of H$\gamma$, along with the lack of other emission lines, makes it difficult to reach a firm conclusion. Since the photospheric temperature is still significant (\s 5000 K), dust formation in the optically thick ejecta is not viable.

There are multiple SNe IIn where dust formation and pre-existing dust are proposed. SN 2010jl is one of the prototypical IIn with multi-wavelength observations, with studies proposing the presence of both pre-existing and newly formed dust \citep{2020ApJ...894..111B,2021ApJ...917...84D}. Other interesting and well-known SNe IIn include the long-lived IIn, SN 2005ip, with a plateau lasting 5 years, followed by a decline \citep{2009ApJ...695.1334S}, accompanied by pre-existing dust \citep{2009ApJ...691..650F}. Recent observations have also revealed active dust formation \citep{2025ApJ...985..262S}. SN 2006jc is a Ibn with evidence of dust formation by t \s 50 \citep{2007ApJ...657L.105F,2009MNRAS.392..894A}, showing marked asymmetry in He I $\lambda$7065 line profile. Similar to these IIn SNe, in SN 2019vxm as well, we are unable to distinguish the origin of the dust or whether there is an interplay between pre-existing and newly formed dust reflected in the photometric and spectroscopic observations.

\section{Summary}
SN 2019vxm is a slow-rising, slow-declining type IIn SN, with an absolute \textit{R}-band magnitude of -20.1 mag, making it brighter than the median magnitude of such objects but less bright than SLSNe. We do not detect any enhanced mass loss over the 4 years preceding the explosion. The photometric analysis indicates that SN 2019vxm evolved through three major phases: the rise, dominated by UV; a post-maximum shallow decline, with a major contribution from the optical; and a steeper IR-dominated decline. The initial evolution is strongly dominated by interaction, while the shallow decline is mainly influenced by the broad component from the interaction. This phase also shows an asymmetric line profile with prominent wings that remain significantly unchanged until 400 days, which we attribute to scattering by an optically thick medium or dust. The degree of scattering remains fairly the same as does the Balmer decrement (\s 3-4), indicating an unchanging scattering region. Hints of ejecta are weak, and blended emission lines of Ca NIR are also observed at this phase. After t \s 400 days, the SN transitions to an IR-dominated phase, where the Balmer line profiles show a rapidly rising degree of asymmetry, with the red flux completely suppressed except for faint wings. The trend toward asymmetry is also reflected in the Balmer decrement, which rises to \s 10. This indicates a very weak H$\beta$ line as compared to H$\alpha$. The line profiles are also blueshifted as they transition from an optical- to an IR-dominated phase. However, H lines at longer and shorter wavelengths will give more insight into the degree of line shift, asymmetry, and wavelength-dependent extinction. Nevertheless, the combination of optical decline and IR rise, increased asymmetry due to the suppression of redder fluxes, and blueshifted line emission is a self-evident signature of dust at play, either pre-existing or newly formed, both of which cannot be ruled out with the available data. However, a high photospheric temperature implies that the inner ejecta is not the site of dust formation at this epoch, since the photospheric temperature is well above the dust formation temperature. We also observe no nebular emission lines at later phases, indicating that the ejecta remained dense at the last epoch or was completely shrouded by the ongoing interaction and/or dust.

\begin{acknowledgments}
We thank the staff of IAO, Hanle, and CREST, Hosakote, who made these observations possible. The facilities at IAO and CREST are operated by the Indian Institute of Astrophysics, Bangalore. 

GCA thanks the Indian National Science Academy for support under the INSA Senior Scientist Programme.

This research has also made use of the NASA/IPAC Extragalactic Database (NED\footnote{\url{https://ned.ipac.caltech.edu}}), which is funded by the National Aeronautics and Space Administration and operated by the California Institute of Technology. This work has used data from the Asteroid Terrestrial-impact Last Alert System (ATLAS) project. The Asteroid Terrestrial-impact Last Alert System (ATLAS) project is primarily funded to search for near-earth asteroids through NASA grants NN12AR55G, 80NSSC18K0284, and 80NSSC18K1575; byproducts of the NEO search include images and catalogues from the survey area. This work was partially funded by Kepler/K2 grant J1944/80NSSC19K0112 and HST GO-15889, and STFC grants ST/T000198/1 and ST/S006109/1. The ATLAS science products have been made possible through the contributions of the University of Hawaii Institute for Astronomy, the Queen's University Belfast, the Space Telescope Science Institute, the South African Astronomical Observatory, and The Millennium Institute of Astrophysics (MAS), Chile.

We also acknowledge Wiezmann Interactive Supernova data REPository\footnote{\url{https://wiserep.weizmann.ac.il}} \citep[(WISeREP)]{2012yaron} for spectral data downloads.

\end{acknowledgments}
\facilities{HCT, Kanata, ATLAS, and Swift}


\newpage

\appendix
\section{MOSFiT models and posteriors}  \label{mosfit_apend}

\begin{figure}[hbt!]
    \centering
    \includegraphics[width=0.8\linewidth]{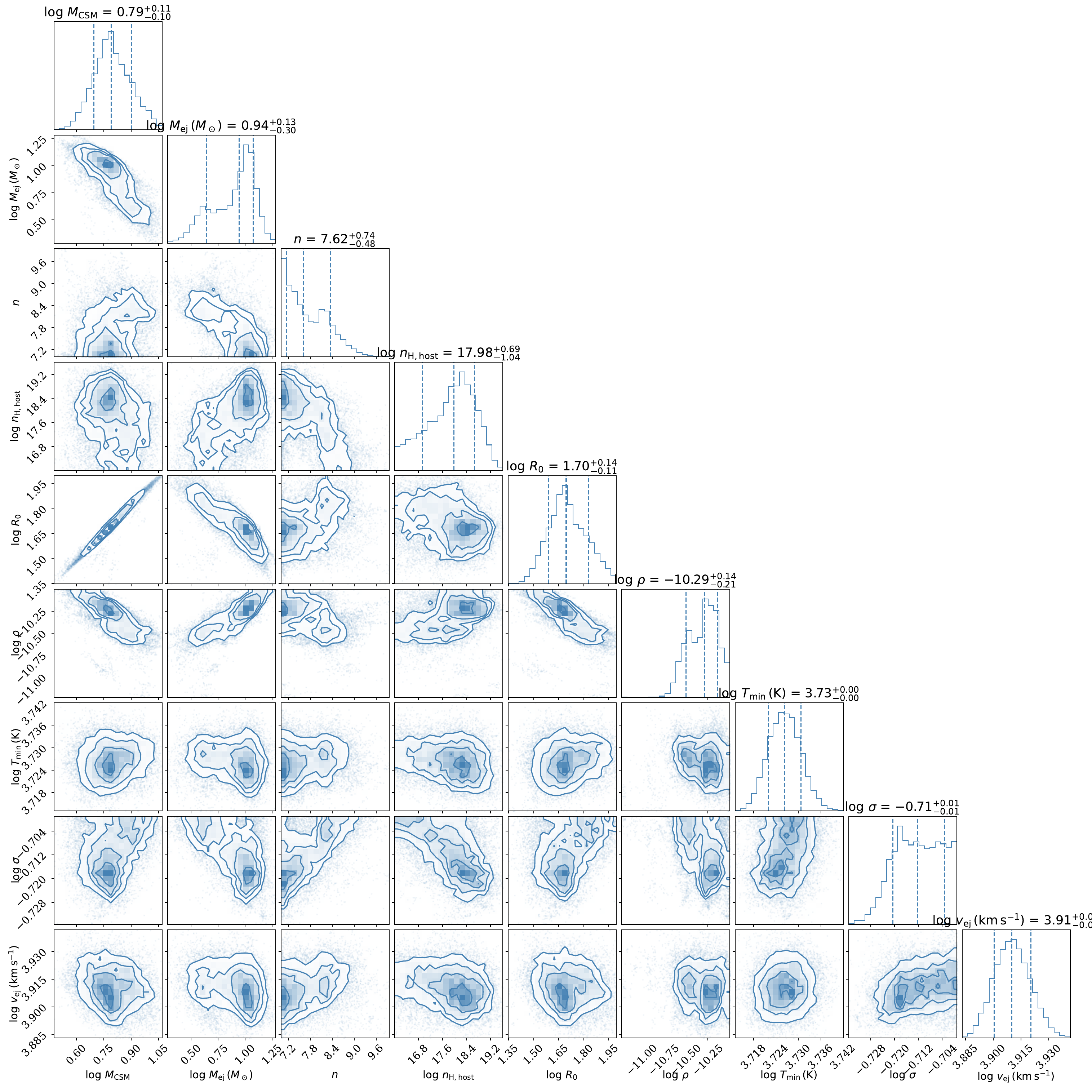}
     \caption{Corner plot till t\s 400 days for s=2.0}
    \label{corner_1.44} 
\end{figure}

\begin{figure}
    \centering
    
    \includegraphics[width=0.5\linewidth]{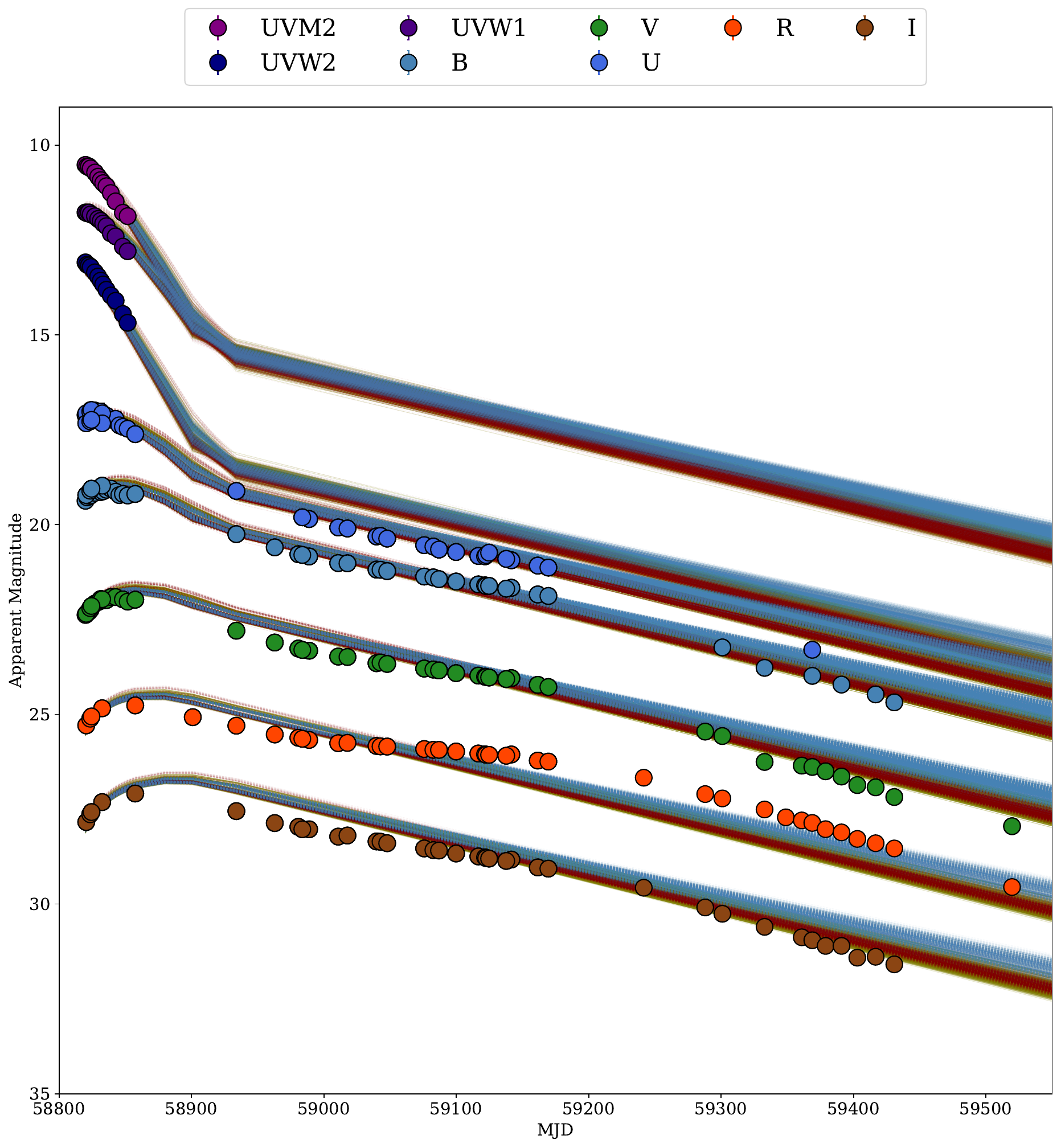}
    \caption{LC obtained by MOSFiT csm model for s=0 (red),s=2 (blue), and s= 1.4 (green)}
    \label{mosfit_s} 
\end{figure}

\begin{figure}
    \centering
    \includegraphics[width=0.6\linewidth]{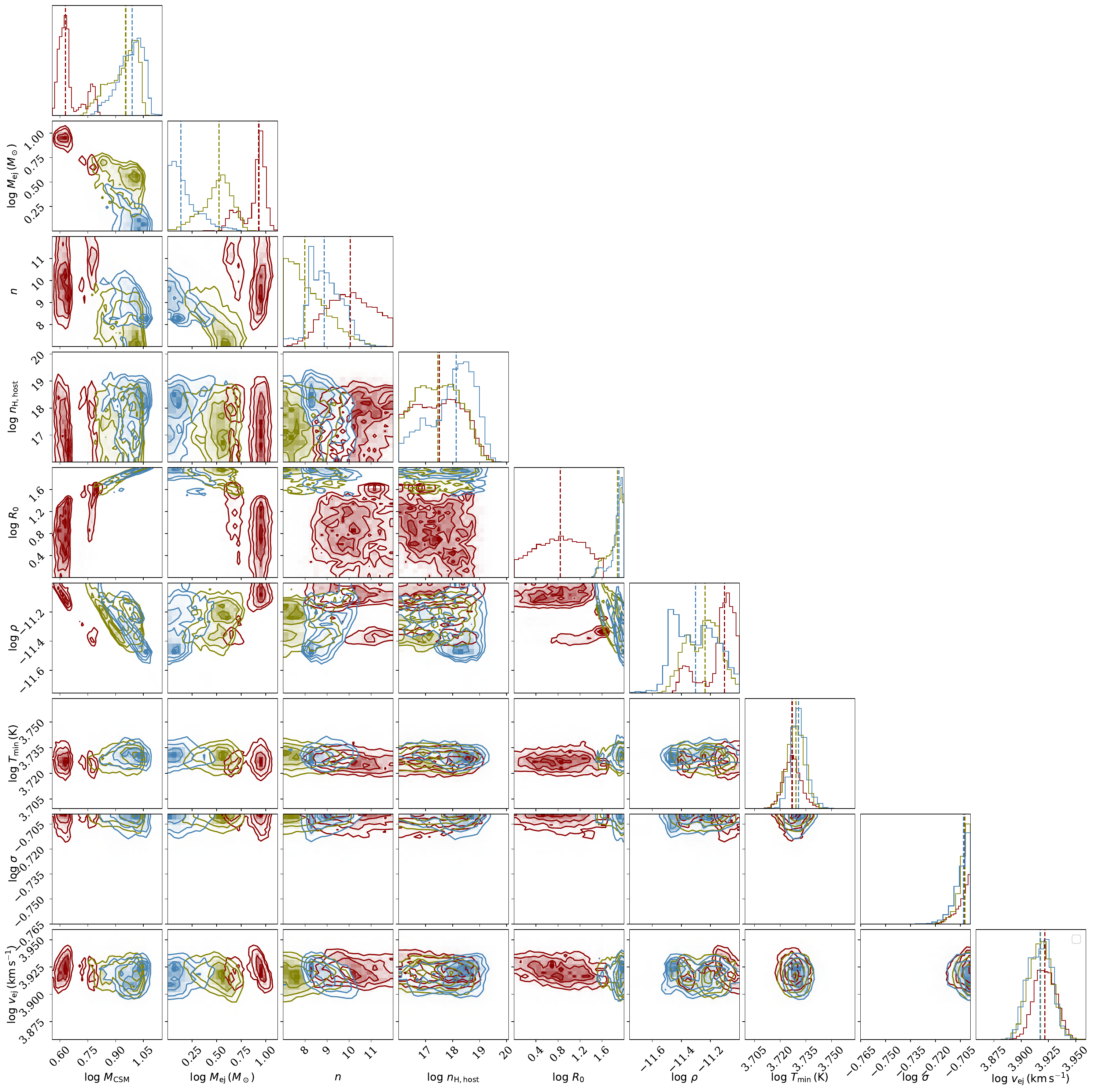}
    \caption{Posteriors obtained by MOSFiT csm model for s=0 (red),s=2 (blue), s=1.4 (green)}
    \label{mosfit_s} 
\end{figure}

\newpage
\newpage

\section{H alpha velocity} \label{halpha_max}
\input{velocity}

\onecolumngrid
\section{Data} \label{append}

\input{HCT_data}
\input{JHK_data}
\input{UVOT_data}
\bibliography{main_bib}{}
\bibliographystyle{aasjournalv7}

\end{document}

%% file: velocity.tex
\begin{longtable}[htb]{ccc}
\caption{The maximum velocity on H-alpha profile in the blue and red side}
\label{vel}\\
\hline
Phase$^{*}$  & \multicolumn{2}{c}{v$_{max}$ (\kms) $^{\dagger}$}  \\
(days)  &  Blue side & Red side\\
\hline \hline
16.6 & 4750 & - \\
19.6 & 5000 &  - \\
20.5 & 4200 &  - \\
28.6 & 4200 &  - \\
53.6 & 9000 &  - \\
129.9 &  9000  &  - \\ 
182.9 & 9000 &  - \\
185.9 & 8800  & 8000 \\
206.8 & 8500 &  9000 \\
213.8 & 8600 &  9000 \\
235.8 & 8200 &  8400 \\
238.8 & 8200  & 8000 \\
243.8 & 8200  & 8000 \\
271.8 & 8200 &  8000 \\
278.7 & 8200 &  8000 \\
301.8 & 7200 &  8200 \\
306.6 & 7800 &  8000 \\
312.7 & 8000 &  8000 \\
318.6 & 7800 &  8200 \\
321.6 & 8000  & 8200 \\
333.7 & 8100 &  8000 \\
337.6 & 8600  & 8000 \\
362.6 & 8400  & 8200 \\
365.6 & 8000  & 8200 \\
372.6 & 8600  & 8000 \\
377.6 & 8600  & 8000 \\
440.6 & 7600  & 8000 \\
489.9 & 7800  & 8000 \\
496.9 & 7300  & 7800 \\
528.9 & 6800 &  7400 \\
555.9 & 6200  & 6000 \\
627.8 & 5600  & 6000 \\
628.6 & 5800  & 5800 \\
634.6 & 6000  & 6000 \\
636.7 & 6200  & 6200 \\
651.8 & 6000  & 5600 \\
672.2 & 5600  & 6000 \\
\hline
\multicolumn{3}{l}{\scriptsize$^*$ Phases are relative to explosion epoch.}\\
\multicolumn{3}{l}{\scriptsize $\dagger$  Error $\sim 780 \rm\ km\ s^{-1}$ (vel step + resolution)}%
\end{longtable}



%% file: HCT_data.tex
\begin{longtable}[htbp]{@{\extracolsep{0.1em}}ccccccc}
\caption{Photometric Observation from HCT (Vega magnitude)}
\label{HCT_data} \\
\hline
\colhead{JD} & \colhead{Phase}$^{*}$ & \colhead{U} &  \colhead{B} & \colhead{V} &  \colhead{R} &  \colhead{I}\\
\hline
2458821.05 & 16.52 & $14.26 \pm 0.08$ & $15.07 \pm 0.03$ & $15.11 \pm 0.01$ & $15.00 \pm 0.02$ & $14.98 \pm 0.03$ \\
2458824.04 & 19.51 & $14.21 \pm 0.05$ & $14.96 \pm 0.03$ & $14.96 \pm 0.01$ & $14.83 \pm 0.02$ & $14.78 \pm 0.23$ \\
2458825.02 & 20.49 & $14.17 \pm 0.05$ & $14.90 \pm 0.02$ & $14.91 \pm 0.01$ & $14.77 \pm 0.02$ & $14.72 \pm 0.02$ \\
2458833.06 & 28.53 & $14.26 \pm 0.22$ & $14.82 \pm 0.02$ & $14.73 \pm 0.02$ & $14.55 \pm 0.02$ & $14.45 \pm 0.03$ \\
2458858.05 & 53.52 & $14.55 \pm 0.17$ & $15.04 \pm 0.05$ & $14.74 \pm 0.02$ & $14.48 \pm 0.03$ & $14.23 \pm 0.04$ \\
2458901.48 & 96.95 & - & - & - & $14.79 \pm 0.04$ & - \\
2458934.42 & 129.89 & $16.04 \pm 0.09$ & $16.10 \pm 0.03$ & $15.56 \pm 0.02$ & $15.01 \pm 0.03$ & $14.69 \pm 0.03$ \\
2458963.40 & 158.87 & - & $16.45 \pm 0.03$ & $15.87 \pm 0.02$ & $15.24 \pm 0.03$ & $15.01 \pm 0.07$ \\
2458981.25 & 176.72 & - & $16.63 \pm 0.03$ & $16.03 \pm 0.01$ & $15.33 \pm 0.02$ & $15.11 \pm 0.02$ \\
2458989.38 & 184.85 & $16.78 \pm 0.08$ & $16.69 \pm 0.03$ & $16.08 \pm 0.02$ & $15.39 \pm 0.09$ & $15.17 \pm 0.09$ \\
2459011.34 & 206.81 & $17.00 \pm 0.09$ & $16.86 \pm 0.03$ & $16.24 \pm 0.02$ & $15.47 \pm 0.02$ & $15.36 \pm 0.04$ \\
2459018.30 & 213.77 & $17.03 \pm 0.05$ & $16.87 \pm 0.03$ & $16.25 \pm 0.02$ & $15.46 \pm 0.04$ & $15.34 \pm 0.03$ \\
2459040.26 & 235.73 & $17.24 \pm 0.11$ & $17.03 \pm 0.03$ & $16.41 \pm 0.02$ & $15.55 \pm 0.03$ & $15.49 \pm 0.03$ \\
2459043.25 & 238.72 & $17.23 \pm 0.08$ & $17.04 \pm 0.03$ & $16.40 \pm 0.02$ & $15.56 \pm 0.02$ & $15.50 \pm 0.04$ \\
2459048.30 & 243.77 & $17.30 \pm 0.05$ & $17.07 \pm 0.02$ & $16.43 \pm 0.01$ & $15.56 \pm 0.03$ & $15.54 \pm 0.03$ \\
2459076.27 & 271.74 & $17.47 \pm 0.09$ & $17.22 \pm 0.02$ & $16.55 \pm 0.01$ & $15.63 \pm 0.03$ & $15.68 \pm 0.03$ \\
2459083.19 & 278.66 & $17.51 \pm 0.07$ & $17.24 \pm 0.02$ & $16.58 \pm 0.02$ & $15.65 \pm 0.02$ & $15.72 \pm 0.02$ \\
2459087.43 & 282.90 & $17.59 \pm 0.20$ & $17.28 \pm 0.03$ & $16.60 \pm 0.02$ & $15.65 \pm 0.04$ & $15.73 \pm 0.03$ \\
2459100.38 & 295.85 & $17.65 \pm 0.13$ & $17.35 \pm 0.03$ & $16.67 \pm 0.02$ & $15.69 \pm 0.02$ & $15.81 \pm 0.04$ \\
2459117.18 & 312.65 & $17.76 \pm 0.12$ & $17.42 \pm 0.02$ & $16.74 \pm 0.02$ & $15.74 \pm 0.02$ & $15.89 \pm 0.03$ \\
2459122.21 & 317.68 & $17.76 \pm 0.08$ & $17.45 \pm 0.01$ & $16.77 \pm 0.01$ & $15.77 \pm 0.01$ & $15.91 \pm 0.02$ \\
2459123.09 & 318.56 & $17.73 \pm 0.11$ & $17.47 \pm 0.02$ & $16.78 \pm 0.01$ & $15.78 \pm 0.01$ & $15.93 \pm 0.02$ \\
2459125.27 & 320.74 & $17.67 \pm 0.15$ & $17.46 \pm 0.04$ & $16.79 \pm 0.01$ & $15.78 \pm 0.03$ & $15.94 \pm 0.03$ \\
2459142.09 & 337.56 & $17.87 \pm 0.11$ & $17.51 \pm 0.06$ & $16.81 \pm 0.07$ & $15.77 \pm 0.08$ & $15.97 \pm 0.07$ \\
2459138.25 & 333.72 & $17.83 \pm 0.12$ & $17.54 \pm 0.04$ & $16.83 \pm 0.03$ & $15.80 \pm 0.03$ & $16.00 \pm 0.03$ \\
2459162.05 & 357.52 & $18.01 \pm 0.11$ & $17.69 \pm 0.04$ & $16.98 \pm 0.03$ & $15.93 \pm 0.04$ & $16.18 \pm 0.05$ \\
2459170.04 & 365.51 & $18.06 \pm 0.08$ & $17.73 \pm 0.02$ & $17.04 \pm 0.01$ & $15.96 \pm 0.02$ & $16.21 \pm 0.02$ \\
2459242.07 & 437.54 & - & - & - & $16.38 \pm 0.02$ & $16.71 \pm 0.05$ \\
2459288.50 & 483.97 & - & - & $18.22 \pm 0.03$ & $16.81 \pm 0.04$ & $17.23 \pm 0.12$ \\
2459301.44 & 496.91 & - & $19.08 \pm 0.04$ & $18.33 \pm 0.03$ & $16.93 \pm 0.02$ & $17.40 \pm 0.06$ \\
2459303.32 & 498.79 & - & $18.59 \pm 0.08$ & $18.30 \pm 0.06$ & $16.85 \pm 0.04$ & $17.26 \pm 0.05$ \\
2459333.34 & 528.81 & - & $19.62 \pm 0.07$ & $19.02 \pm 0.08$ & $17.22 \pm 0.03$ & $17.74 \pm 0.07$ \\
2459349.40 & 544.87 & - & - & - & $17.43 \pm 0.07$ & - \\
2459361.42 & 556.89 & - & - & $19.11 \pm 0.03$ & $17.51 \pm 0.06$ & $18.02 \pm 0.14$ \\
2459369.36 & 564.83 & $20.23 \pm 0.20$ & $19.83 \pm 0.04$ & $19.15 \pm 0.03$ & $17.58 \pm 0.03$ & $18.10 \pm 0.07$ \\
2459379.40 & 574.87 & - & - & $19.26 \pm 0.04$ & $17.74 \pm 0.03$ & $18.24 \pm 0.06$ \\
2459391.40 & 586.87 & - & $20.06 \pm 0.08$ & $19.40 \pm 0.04$ & $17.82 \pm 0.02$ & $18.24 \pm 0.04$ \\
2459403.38 & 598.85 & - & - & $19.62 \pm 0.04$ & $18.00 \pm 0.03$ & $18.56 \pm 0.04$ \\
2459417.22 & 612.69 & - & $20.32 \pm 0.06$ & $19.68 \pm 0.04$ & $18.11 \pm 0.02$ & $18.53 \pm 0.06$ \\
2459431.31 & 626.78 & - & $20.53 \pm 0.05$ & $19.94 \pm 0.05$ & $18.25 \pm 0.04$ & $18.73 \pm 0.07$ \\
2459520.22 & 715.69 & - & - & $20.71 \pm 0.10$ & $19.26 \pm 0.04$ & - \\
\hline
\multicolumn{7}{l}{{$^*$ Phases are relative to explosion epoch, JD 2458804.53}}\\
\end{longtable}
\raggedbottom

%% file: JHK_data.tex
\begin{longtable}[htbp]{@{\extracolsep{0.1em}}ccccc}
\caption{Photometric Observation from Kanata Telscope (Vega)}
\label{JHK_Data} \\
\hline
\colhead{MJD} & \colhead{Phase}$^{*}$ & \colhead{J} &  \colhead{H} & \colhead{Ks} \\
\hline
58821.500 & 17.470 & $14.94 \pm 0.03$ & - & $14.75 \pm 0.05$ \\
58824.400 & 20.370 & $14.84 \pm 0.02$ & $14.74 \pm 0.08$ & $14.60 \pm 0.10$ \\
58825.500 & 21.470 & $14.68 \pm 0.02$ & $14.63 \pm 0.03$ & $14.49 \pm 0.06$ \\
58827.500 & 23.470 & $14.62 \pm 0.03$ & - & $14.40 \pm 0.05$ \\
58828.500 & 24.470 & $14.62 \pm 0.01$ & $15.18 \pm 0.03$ & $14.43 \pm 0.05$ \\
58829.500 & 25.470 & $14.59 \pm 0.01$ & - & - \\
58830.400 & 26.370 & $14.49 \pm 0.02$ & $14.39 \pm 0.05$ & $14.33 \pm 0.07$ \\
58840.400 & 36.370 & - & $14.03 \pm 0.02$ & - \\
58841.400 & 37.370 & $14.19 \pm 0.05$ & $14.12 \pm 0.13$ & $13.89 \pm 0.06$ \\
58845.400 & 41.370 & $14.13 \pm 0.02$ & $14.05 \pm 0.06$ & $13.85 \pm 0.04$ \\
58851.400 & 47.370 & $14.08 \pm 0.02$ & - & $13.78 \pm 0.02$ \\
58858.400 & 54.370 & $14.02 \pm 0.02$ & $13.78 \pm 0.06$ & $13.71 \pm 0.04$ \\
58866.500 & 62.470 & $13.98 \pm 0.01$ & - & $13.63 \pm 0.03$ \\
58909.800 & 105.770 & $14.10 \pm 0.01$ & $13.82 \pm 0.03$ & $13.50 \pm 0.04$ \\
58925.800 & 121.770 & $14.20 \pm 0.02$ & $13.95 \pm 0.02$ & $13.57 \pm 0.03$ \\
58944.700 & 140.670 & $14.56 \pm 0.01$ & $14.27 \pm 0.05$ & $13.64 \pm 0.07$ \\
58949.800 & 145.770 & $14.37 \pm 0.02$ & $14.09 \pm 0.03$ & $13.68 \pm 0.03$ \\
58967.700 & 163.670 & $14.53 \pm 0.03$ & $14.31 \pm 0.14$ & $13.78 \pm 0.04$ \\
58991.700 & 187.670 & $14.62 \pm 0.02$ & $14.39 \pm 0.05$ & $13.93 \pm 0.06$ \\
58995.800 & 191.770 & $14.68 \pm 0.02$ & $14.38 \pm 0.06$ & $13.96 \pm 0.05$ \\
59007.700 & 203.670 & $14.66 \pm 0.02$ & $14.89 \pm 0.17$ & $13.94 \pm 0.05$ \\
59019.600 & 215.570 & - & $14.54 \pm 0.04$ & $14.16 \pm 0.04$ \\
59021.600 & 217.570 & $14.81 \pm 0.03$ & $14.51 \pm 0.04$ & $13.98 \pm 0.05$ \\
59023.600 & 219.570 & $14.80 \pm 0.02$ & $14.64 \pm 0.04$ & $14.04 \pm 0.09$ \\
59031.700 & 227.670 & $14.91 \pm 0.02$ & $14.59 \pm 0.05$ & $14.04 \pm 0.05$ \\
59060.800 & 256.770 & $14.99 \pm 0.04$ & $14.83 \pm 0.14$ & $14.06 \pm 0.18$ \\
59061.500 & 257.470 & $14.88 \pm 0.06$ & - & - \\
59066.600 & 262.570 & $14.96 \pm 0.02$ & $14.75 \pm 0.03$ & - \\
59093.600 & 289.570 & $15.10 \pm 0.04$ & $14.91 \pm 0.05$ & - \\
59152.600 & 348.570 & $15.32 \pm 0.02$ & - & $14.44 \pm 0.06$ \\
59163.500 & 359.470 & $15.39 \pm 0.02$ & - & $14.39 \pm 0.07$ \\
59176.500 & 372.470 & $15.42 \pm 0.05$ & - & $14.44 \pm 0.02$ \\
59192.400 & 388.370 & $15.53 \pm 0.02$ & $15.17 \pm 0.04$ & $14.30 \pm 0.07$ \\
59204.500 & 400.470 & - & $15.13 \pm 0.20$ & - \\
59287.900 & 483.870 & - & $14.94 \pm 0.03$ & $14.22 \pm 0.03$ \\
59302.800 & 498.770 & $16.04 \pm 0.02$ & $15.05 \pm 0.01$ & - \\
59325.700 & 521.670 & $16.08 \pm 0.01$ & - & $14.14 \pm 0.02$ \\
59343.800 & 539.770 & $16.36 \pm 0.01$ & - & $14.04 \pm 0.01$ \\
59365.700 & 561.670 & $16.29 \pm 0.01$ & - & $14.04 \pm 0.01$ \\
59367.700 & 563.670 & $16.48 \pm 0.01$ & $15.34 \pm 0.01$ & $14.12 \pm 0.01$ \\
59372.700 & 568.670 & $16.22 \pm 0.01$ & - & - \\
59384.700 & 580.670 & $16.60 \pm 0.02$ & - & - \\
59386.700 & 582.670 & $16.42 \pm 0.02$ & $15.17 \pm 0.02$ & - \\
59414.700 & 610.670 & $16.56 \pm 0.01$ & $15.49 \pm 0.01$ & $14.20 \pm 0.01$ \\
59423.500 & 619.470 & $17.14 \pm 0.02$ & $15.59 \pm 0.01$ & - \\
59454.600 & 650.570 & $16.84 \pm 0.03$ & $15.37 \pm 0.02$ & $14.06 \pm 0.02$ \\
59675.800 & 871.770 & - & $16.68 \pm 0.04$ & $15.22 \pm 0.03$ \\
\hline
\end{longtable}
\raggedbottom

%% file: UVOT_data.tex
\begin{longtable}[htbp]{@{\extracolsep{0.1em}}cccccccc}
\caption{Photometric Observation from Swift/UVOT (Vega)}
\label{UVOT_Data} \\
\hline
\colhead{JD} & \colhead{Phase}$^{*}$ & \colhead{UVW2} &  \colhead{UVM2} & \colhead{UVW1} & \colhead{UVU} & \colhead{UVB} & \colhead{UBV} \\
\hline
2458820.500 & 15.970 & $13.95 \pm 0.03$ & $13.88 \pm 0.02$ & $13.89 \pm 0.03$ & $14.06 \pm 0.03$ & $15.24 \pm 0.03$ & $15.17 \pm 0.05$ \\
2458821.700 & 17.170 & $14.01 \pm 0.03$ & $13.92 \pm 0.03$ & $13.91 \pm 0.03$ & $14.05 \pm 0.04$ & $15.15 \pm 0.04$ & $15.12 \pm 0.06$ \\
2458823.000 & 18.470 & $14.03 \pm 0.03$ & $13.92 \pm 0.03$ & $13.89 \pm 0.03$ & $13.99 \pm 0.03$ & $15.11 \pm 0.03$ & $15.09 \pm 0.04$ \\
2458824.400 & 19.870 & $14.07 \pm 0.03$ & $13.97 \pm 0.03$ & $13.94 \pm 0.03$ & $13.93 \pm 0.03$ & $15.10 \pm 0.03$ & $15.03 \pm 0.04$ \\
2458827.600 & 23.070 & $14.21 \pm 0.03$ & $14.08 \pm 0.03$ & $13.99 \pm 0.03$ & $13.94 \pm 0.03$ & $15.03 \pm 0.03$ & $14.87 \pm 0.04$ \\
2458830.000 & 25.470 & $14.32 \pm 0.03$ & $14.19 \pm 0.03$ & $14.06 \pm 0.03$ & $13.98 \pm 0.03$ & $14.95 \pm 0.03$ & $14.83 \pm 0.05$ \\
2458832.000 & 27.470 & $14.43 \pm 0.03$ & $14.28 \pm 0.03$ & $14.11 \pm 0.03$ & $13.97 \pm 0.03$ & $15.01 \pm 0.03$ & $14.75 \pm 0.04$ \\
2458833.800 & 29.270 & $14.53 \pm 0.03$ & $14.36 \pm 0.03$ & $14.18 \pm 0.03$ & $14.03 \pm 0.03$ & $14.99 \pm 0.03$ & $14.79 \pm 0.04$ \\
2458836.300 & 31.770 & $14.67 \pm 0.03$ & $14.44 \pm 0.03$ & $14.25 \pm 0.03$ & $14.08 \pm 0.04$ & $14.95 \pm 0.04$ & $14.78 \pm 0.05$ \\
2458839.600 & 35.070 & $14.82 \pm 0.03$ & $14.62 \pm 0.03$ & $14.44 \pm 0.03$ & $14.18 \pm 0.03$ & $14.92 \pm 0.03$ & $14.70 \pm 0.04$ \\
2458843.200 & 38.670 & $14.96 \pm 0.03$ & $14.84 \pm 0.03$ & $14.52 \pm 0.03$ & $14.16 \pm 0.04$ & $15.00 \pm 0.03$ & $14.70 \pm 0.05$ \\
2458846.000 & 41.470 & - & - & - & $14.33 \pm 0.03$ & $15.09 \pm 0.03$ & - \\
2458848.700 & 44.170 & $15.31 \pm 0.03$ & $15.14 \pm 0.03$ & $14.79 \pm 0.03$ & $14.37 \pm 0.03$ & $15.06 \pm 0.03$ & $14.76 \pm 0.04$ \\
2458852.300 & 47.770 & $15.54 \pm 0.04$ & $15.23 \pm 0.04$ & $14.91 \pm 0.04$ & $14.42 \pm 0.04$ & $15.10 \pm 0.04$ & $14.82 \pm 0.05$ \\

\hline
\end{longtable}
\raggedbottom